\documentclass[a4paper,fleqn]{cas-sc}

\usepackage[numbers,sort&compress]{natbib}
\usepackage{amsmath,amssymb,amsfonts}
\usepackage{algorithm}
\usepackage{algorithmic}
\usepackage{graphicx}
\usepackage{textcomp}
\usepackage{xcolor}
\usepackage{caption}
\usepackage{subcaption}

\usepackage{colortbl}
\usepackage{dhucs} 
\usepackage{booktabs} 
\usepackage{multirow}
\usepackage{comment}
\usepackage{lineno}
\usepackage{color} 
\usepackage{setspace} 
\usepackage{graphicx} 
\usepackage{makecell}
\usepackage{enumitem} 
\usepackage{relsize}
\usepackage{tabularx}
\usepackage{diagbox}
\usepackage{amssymb}
\usepackage{xcolor}
\usepackage{colortbl}
\usepackage{hhline}
\usepackage{makecell}

\usepackage{array}

\newcommand\refFigure[1]{Fig.~\ref{#1}}
\newcommand\refTable[1]{Table~\ref{#1}}

\newcommand\refAlgo[1]{Algorithm~\ref{#1}}

\newcommand\blind[1]{XXXX}

\begin{document}
\let\WriteBookmarks\relax
\def\floatpagepagefraction{1}
\def\textpagefraction{.001}

\shorttitle{}    


\title [mode = title]{{Asymptotic Error Bounds and Fractional-Bit Design for Fixed-Point Grover’s Quantum Algorithm Emulation}}  



\author[]{Seonghyun Choi}
\author[]{Kyeongwon Lee}
\author[]{Jongin Choi}
\author[]{Woojoo Lee}
\ead{space@cau.ac.kr}
\cormark[1]

\affiliation[]{organization={Chung-Ang University},
    city={Seoul},
    postcode={06974}, 
    country={Korea}}
    \cortext[cor1]{Corresponding author}


\begin{abstract}
Quantum computing (QC) emulators, which simulate quantum algorithms on classical hardware, serve as indispensable platforms for designing and testing quantum algorithms prior to the availability of scalable quantum computers. A critical challenge in classical hardware-based QC emulation is managing numerical errors arising from finite arithmetic precision, particularly truncation errors in resource-efficient fixed-point arithmetic. Despite its importance, systematic studies addressing how fixed-point truncation errors quantitatively impact the accuracy of quantum algorithms are limited.
In this paper, we develop a rigorous quantitative framework to analyze how truncation errors propagate and accumulate in fixed-point QC emulation, with a particular focus on Grover’s quantum search algorithm. 
First, we introduce a simplified two-value amplitude representation for quantum states during Grover’s iterations and provide a theoretical proof validating its correctness. Leveraging this representation, we derive explicit mathematical expressions that characterize the accumulation of truncation errors across multiple quantum gate operations. We then quantify the overall emulation error using the $\ell_2$ distance between the ideal and emulated probability distributions, obtaining explicit asymptotic bounds that scale as $O(2^{n-f})$, where $n$ denotes the number of qubits and $f$ is the fractional-bit precision.
Our theoretical model is validated through extensive numerical simulations and empirical experiments conducted on a practical fixed-point QC emulator. The results confirm that the observed errors precisely align with our theoretical predictions and exhibit predictable exponential scaling behavior. Finally, we provide a practical closed-form formula to determine the minimal fractional-bit precision required to meet a specified error threshold, offering clear and actionable guidelines for emulator designers to optimize accuracy versus resource utilization.

\end{abstract}




\maketitle

\section{Introduction}
\label{sec:intro}

Over the past few decades, \emph{quantum computing} (QC) has been vigorously studied in both academia~\cite{Kim:Nature23,Kikuchi:Nature23} and industry~\cite{Google:Science20,Zhang:Nature24,Microsoft:Nature25}. 
By leveraging quantum phenomena such as superposition and entanglement, quantum computers operate on fundamentally different computational principles compared to classical machines. 
Owing to this unique mode of operation, quantum computers can dramatically reduce the computational overhead for certain problems and potentially solve instances that are intractable on classical supercomputers~\cite{Ladd:Nature10,Cao:ACS19,Gyongyosi:Elsevier19}. 
In fact, multiple experiments have experimentally demonstrated the superiority of quantum hardware over classical counterparts in carefully chosen tasks~\cite{Arute:Nature19,Zhong:Science20}.  

To translate these performance advantages into real-world applications, considerable effort has been dedicated to developing QC hardware via various physical implementations~\cite{Kelly:Nature15,Maurand:Nature16,Friis:APS18}. 
In parallel, researchers are exploring quantum algorithms that aim to efficiently tackle practical problems, such as optimization~\cite{Ajagekar:Elsevier19,Zhou:PRX20,Wang:ASME23}, machine learning~\cite{Cai:APS15,Ramezani:IEEE20,Lloyd:arxiv13}, and complex system simulations~\cite{Iriyama:AMC12,Ollitrault:ACS21,Outeiral:wiley21}---all of which have the potential to surpass classical methods in terms of computational throughput or memory scaling.

Despite these advances, current quantum hardware continues to face significant challenges in \emph{scalability} and \emph{error correction}, leading to hardware that is not yet at a practical, large-scale “quantum network” level. 
Because of these limitations, researchers often lack direct hardware access to develop, test, and evaluate sophisticated quantum algorithms. 
To bridge this gap, \emph{QC emulators}---which simulate or emulate quantum circuits on classical machines---have been actively investigated~\cite{Buluta:Science09,Haner:IEEE16, Li:IEEE21,Altman:PRX21,Choi:AIMS24,Choi:arXiv25}. 
These emulators fill a critical role in quantum computing research, enabling rapid prototyping and testing of quantum algorithms before fully scalable quantum hardware becomes available.

A QC emulator models quantum states and gates as vectors and matrices, respectively, on a classical computer~\cite{Khalid:IEEE04}. However, numerical errors originating from the {classical} hardware---such as floating-point or fixed-point inaccuracies---can directly degrade the fidelity of the emulated quantum operations. 
Identifying and quantifying these classical errors are therefore essential for designing reliable QC emulators. 
Although a few works have discussed precision issues in QC simulation, {systematic} investigations that examine how the choice of arithmetic precision directly impacts the \emph{accuracy} of specific quantum algorithms remain sparse. 
In particular, there is a lack of focused studies that assess the interplay between classical arithmetic errors and emulator accuracy when a target quantum algorithm is fixed.

In this paper, we address this gap by proposing a \emph{quantitative framework} to analyze how classical arithmetic errors influence the accuracy of a \emph{fixed-point-based} QC emulator. From this analysis, we derive a closed-form formula that computes the \emph{optimal precision} (i.e., the number of fractional bits) required to keep errors under a specified target---thus providing practical guidelines for resource optimization.

To concretize our study, we focus on \emph{Grover’s algorithm}~\cite{Grover:ACM96}, a quantum search procedure widely regarded for its quadratic speedup in unstructured search. This speedup benefits numerous application domains demanding efficient data retrieval, and emulators specialized to Grover’s algorithm have indeed been proposed~\cite{Bag:IEEE22,Choi:AIMS24}. Grover’s algorithm exploits quantum superposition to simultaneously probe multiple data items, requiring only $O(\sqrt{N})$ queries for a dataset of size~$N$, whereas classical algorithms need $O(N)$~\cite{Bennett:SIAM97,Boyer:Wiley98,Zalka:PRA99}. This offers a \emph{quadratic} improvement over the classical approach, empowering quantum computers to address search problems at scales that would be infeasible for classical machines. The algorithm also underpins potential breakthroughs in cryptography~\cite{Preston:IEEE22,Grassl:Springer16,Schwabe:Springer16} and optimization~\cite{Habibi:MDPI22,Bogatyrev:IEEE23application,Chakrabarty:Springer17}.

Next, we adopt a {fixed-point arithmetic} environment for the QC emulator, reflecting its predominant usage in hardware-based QC simulations due to resource efficiency and implementational simplicity. 
In particular, we investigate how the resulting \emph{truncation errors}---arising from representing real numbers with a finite number of fractional bits---accumulate over multiple gate operations. 
Although fixed-point arithmetic mitigates exponential resource overheads more effectively than floating-point in many practical designs, the finite precision of fractional bits inevitably leads to truncation. 
In this work, we pinpoint the specific operations in Grover’s algorithm where these truncations occur and rigorously quantify how they propagate through consecutive computational steps.

Our approach begins by proposing a simplified representation of the quantum state within Grover’s algorithm and providing a theoretical proof of its validity. 
Building on this representation, we derive a mathematical formula for the $\ell_2$ distance between the true and emulated measurement probability distributions under varying qubit counts and fractional-bit precision. 
We then investigate how this $\ell_2$ distance scales with these parameters, forming a rigorous theoretical foundation for understanding and predicting emulator errors. 
We demonstrate the soundness of our derivation by comparing the theoretical error bound with actual measurements from a working fixed-point QC emulator specialized for Grover’s algorithm (henceforth referred to as the \textbf{FP-Grover} emulator). 
Finally, we leverage our formula to propose the {minimum required fractional-bit precision} that keeps emulator errors below a target threshold---a practical guideline for designing {FP-Grover emulators that balance accuracy and resource constraints.

The key contributions of this paper are summarized as follows:
\begin{enumerate}
    \item \textbf{Numerical Error Analysis}:
    We present a systematic method to analyze the truncation errors in a FP-Grover emulator, deriving how these errors manifest in the measurement probability distribution (measured by the $\ell_2$ distance).
    
    \item \textbf{Empirical Validation}:
    We evaluate how the number of qubits and fractional bits affect the overall accuracy loss. Through comparisons of theoretical predictions and emulator outputs, we verify the validity of our approach.
    
    \item \textbf{Design Guideline}:
    We formulate a closed-form expression relating desired error thresholds to the necessary fixed-point precision. This formula can be used directly by designers to achieve resource-optimized emulators while guaranteeing specified accuracy levels.
\end{enumerate}

This paper is structured in the following way.
Section 2 briefly reviews the essential background on quantum states, gates, and the principles of Grover’s algorithm.
Section 3 details our simplified state representation technique for Grover’s algorithm and derives the error accumulation formulas under fixed-point arithmetic.
Section 4 compares the proposed theoretical formula with measured error values from a practical emulator and discusses how to determine the minimum required fractional bits for a given error target.
Finally, Section 5 offers concluding remarks and future perspectives.

\section{Fixed-Point Grover Emulation: A Preliminary}
\label{sec:prelim}
In this preliminary section, we briefly introduce the fundamental quantum computing concepts and Grover’s algorithm necessary to understand the fixed-point arithmetic emulation and associated truncation errors analyzed subsequently.

\subsection{Quantum Information and Gate Operations}

In classical computing, information is stored in an array of bits, and logical gates process these bits to perform computations. 
In quantum computing, qubits and quantum gates respectively serve as counterparts to classical bits and logical gates. 
Specifically, the amplitude and phase of a qubit encode information, while physical operations that modify these amplitudes and phases drive the computation. 
Unlike classical bits, which can only assume one of two discrete values (0 or 1), a qubit can reside in either of the basis states $\lvert 0 \rangle$, $\lvert 1 \rangle$, or any superposition thereof. 
This superposition property is rooted in uniquely quantum phenomena, giving rise to the enhanced computational power that quantum computers can exhibit relative to classical devices.

A single-qubit system has two basis states, each represented by a unit column vector:
\begin{equation}
\lvert 0 \rangle \;=\; \begin{bmatrix} 1 \\ 0 \end{bmatrix}, 
\quad
\lvert 1 \rangle \;=\; \begin{bmatrix} 0 \\ 1 \end{bmatrix}.
\end{equation}
Any superposition of $\lvert 0 \rangle$ and $\lvert 1 \rangle$ can be written as:
\begin{equation}
\label{eq:state_single_qb}
\lvert \psi \rangle \;=\; \alpha \lvert 0 \rangle \;+\; \beta \lvert 1 \rangle
\;=\;
\begin{bmatrix} \alpha \\ \beta \end{bmatrix},
\end{equation}
where $\alpha$ and $\beta$ are complex amplitudes that satisfy $\|\alpha\|^2 + \|\beta\|^2 = 1$. Hence, $\lvert \psi \rangle$ is a unit vector.

A quantum system comprising $n$ qubits spans $2^n$ possible basis states. Labeling these basis states as $\lvert 0 \rangle, \lvert 1 \rangle, \ldots, \lvert 2^n - 1 \rangle$, we can model the state of the system by a $2^n$-dimensional column vector:
\begin{equation}
\label{eq:state_multiple_qb}
\lvert \psi \rangle
\;=\;
\sum_{i=0}^{2^n - 1} \alpha_i \,\lvert i \rangle
\;=\;
\begin{bmatrix}
\alpha_0 \;\; \alpha_1 \;\; \ldots \;\; \alpha_{2^n - 1}
\end{bmatrix}^{\!T},
\end{equation}
where each $\alpha_i$ is a complex number. As in the single-qubit case, these amplitudes must satisfy
$\sqrt{\sum_{i=0}^{2^n-1} \|\alpha_i\|^2} = 1$ to ensure $\lvert \psi \rangle$ is a unit vector.

When measuring a multi-qubit state in the computational basis, the system probabilistically collapses to one of the basis states. Specifically, the probability of obtaining $\lvert i \rangle$ upon measurement is given by
\begin{equation}
\label{eq:prob_is_square}
P\bigl(\lvert \psi \rangle \rightarrow \lvert i \rangle\bigr)
\;=\;
\|\alpha_i\|^2,
\end{equation}
and the sum of all measurement probabilities is $1$, implying
\begin{equation}
\label{eq:uclead_norm_square_sum}
\|\lvert \psi \rangle\|
\;=\;
\sqrt{\sum_{i=0}^{2^n-1} \|\alpha_i\|^2}
\;=\;
1.
\end{equation}

\vspace{0.3em}
\noindent
\paragraph*{Quantum Gates.}\,
A quantum gate can be represented by a unitary matrix that acts on the state vector. Because quantum states must remain normalized (as in Eq.~\eqref{eq:uclead_norm_square_sum}), the operator must preserve the vector norm, making it unitary. Table~\ref{table:quantum_gates} lists several commonly used single- and two-qubit gates, along with their matrix forms. Multi-qubit gates can be constructed from these basic operations via tensor (Kronecker) products. For instance, if $U_1, U_2, \ldots, U_n$ are single-qubit gates applied in parallel to the $n$ qubits, the overall $n$-qubit operation $U$ is the Kronecker product:
\begin{equation}
\label{eq:n_qubit_op}
U 
\;=\;
U_1 \;\otimes\; U_2 \;\otimes\;\cdots\;\otimes\; U_n.
\end{equation}
\begin{table}[t]
\centering
\caption{Representative quantum gates for single-qubit and two-qubit systems.}
\label{table:quantum_gates}
\begin{tabular}{|c|c|c||c|c|c|}
\hline
Gate & Symbol & Unitary Matrix & Gate & Symbol & Unitary Matrix \\ \hline
\multirow{2}{*}{\makecell{\vspace{5pt}\\Hadamard}} & \multirow{2}{*}{\makecell{\vspace{5pt}\\$H$}} 
   & \multirow{2}{*}{\makecell{\vspace{-2pt} \\
     $\displaystyle
     \frac{1}{\sqrt{2}}
     \begin{bmatrix}
       1 & 1 \\
       1 & -1 
     \end{bmatrix}
     $
   }}
   & \multirow{2}{*}{\makecell{\vspace{5pt}\\Phase-shift}} & $P$ &
     \makecell{\vspace{-7pt} \\
     $\displaystyle
     \begin{bmatrix}
       1 & 0 \\
       0 & i 
     \end{bmatrix}$ 
     \\ \vspace{-7pt}
     }
\\\cline{5-6}
 & & & & $T$ &
 \makecell{\vspace{-7pt} \\
   $\displaystyle
   \begin{bmatrix}
     1 & 0 \\
     0 & e^{\,i\pi/4}
   \end{bmatrix}$
   \\ \vspace{-7pt}
   }
\\ \hline
\multirow{3}{*}{\makecell{\vspace{18pt}\\Pauli}} & $X$ &
    \makecell{\vspace{-7pt}\\
   $\displaystyle
   \begin{bmatrix}
     0 & 1 \\
     1 & 0 
   \end{bmatrix}$
   \\ \vspace{-7pt} 
   }
   & \multirow{3}{*}{\makecell{\vspace{18pt}\\Controlled-NOT}} & \multirow{3}{*}{\makecell{\vspace{18pt}\\$CNOT$}} & \multirow{3}{*}{\makecell{\vspace{0pt}\\
    \hspace{1pt}
    $\displaystyle
     \begin{bmatrix} 
       1 & 0 & 0 & 0 \\
       0 & 1 & 0 & 0 \\
       0 & 0 & 0 & 1 \\
       0 & 0 & 1 & 0 
     \end{bmatrix}
     $
     }
   }
\\\cline{2-3}
 & $Y$ &  \makecell{\vspace{-7pt} \\
   $\displaystyle
   \begin{bmatrix}
     0 & -i \\
     i & \;1 
   \end{bmatrix}$
   \\ \vspace{-7pt}
   }
   & & & 
\\\cline{2-3}
 & $Z$ &   
    \makecell{\vspace{-7pt} \\
   $\displaystyle
   \begin{bmatrix}
     1 & 0 \\
     0 & -1 
   \end{bmatrix}$
   \\ \vspace{-7pt}
   }
   & & & 
\\ \hline
\end{tabular}
\end{table}

\subsection{Grover’s Algorithm}
\label{subsec:grover_algorithm}

\emph{Grover’s algorithm} is a quantum search procedure that identifies an $x$ satisfying $f(x)=1$, where $f(x)$ is defined as
\begin{equation}
\label{eq:grover_fx}
f(x) \;=\;
\begin{cases}
   1 & \text{if } x \in S,\\
   0 & \text{otherwise},
\end{cases}
\end{equation}
and $S$ is the set of valid solutions (i.e., $s \in S$ implies $f(s)=1$). Applied to an $n$-qubit system, Grover’s algorithm first prepares the uniform superposition over all $2^n$ basis states, then iteratively applies $k$ rounds of a \emph{Grover operator} $G$ to amplify the amplitudes of solution states. 
Denoting the final state by $\lvert \psi_{\mathrm{final}} \rangle$, one can measure it with high probability of collapsing onto a solution. 
Mathematically,
\begin{equation}
\label{eq:grover_and_G}
\lvert \psi_{\mathrm{final}} \rangle \;=\; G^k \,\lvert n \rangle,
\quad
\text{where}
\quad
G \;=\; \bigl(H^{\otimes n} \,Z_{\mathrm{or}}\, H^{\otimes n}\bigr)\,\bigl(Z_f\bigr).
\end{equation}
Here, $\lvert 0^n \rangle$ denotes all $n$ qubits in the $\lvert 0\rangle$ state, and $H^{\otimes n}$ is a Hadamard gate acting on each qubit. The number of iterations $k$ is approximately
\begin{equation}
\label{eq:k}
k 
\;=\; 
\left\lfloor
\frac{\pi}{4}
\sqrt{\frac{2^n}{n_s}}
\right\rceil,
\end{equation}
where $\lfloor \,\cdot\,\rceil$ rounds to the nearest integer and $n_s$ is the number of solutions in $S$. 
The initial uniform superposition
\begin{equation}
\label{eq:uniform_superposition_state}
\lvert n \rangle
\;=\;
H^{\otimes n} \,\lvert 0^n \rangle
\;=\;
\frac{1}{\sqrt{2^n}}
\sum_{i=0}^{2^n - 1} \lvert i \rangle.
\end{equation}

The $G$ operator in Eq.~\eqref{eq:grover_and_G} comprises two components: an \emph{oracle} gate, represented by the matrix $Z_f$, and a \emph{diffusion} gate, given by $H^{\otimes n} Z_{\mathrm{or}} H^{\otimes n}$. As illustrated in the middle of~\refFigure{fig:G_op_graphic}, the oracle gate identifies solution basis states $\lvert s \rangle$ and flips their phase, effectively negating the amplitude coefficient for each solution state. Meanwhile, the diffusion gate, shown on the right in~\refFigure{fig:G_op_graphic}, reflects the amplitude of all $2^n$ basis states about their average. Concretely, the oracle $Z_f$ is defined by
\begin{equation}
\label{eq:oracle_gate}
Z_f \,\lvert x \rangle
\;=\;
(-1)^{\,f(x)} \,\lvert x \rangle
\;=\;
\begin{cases}
   -\,\lvert x \rangle & x \in S, \\
   \phantom{-}\lvert x \rangle & x \notin S,
\end{cases}
\end{equation}
while the diffusion operator $H^{\otimes n}Z_{\mathrm{or}}H^{\otimes n}$ can be written as
\begin{align}
\label{eq:diffusion_gate}
H^{\otimes n} \,Z_{\mathrm{or}}\, H^{\otimes n}
&\;=\;
2\,\lvert n \rangle\langle n \rvert \;-\; I_n
\;=\;
\begin{bmatrix}
   2^{-\,n+1} & \cdots & 2^{-\,n+1}
   \\
   \vdots & \ddots & \vdots
   \\
   2^{-\,n+1} & \cdots & 2^{-\,n+1}
\end{bmatrix}
\;-\;
I_n, 
\\
\label{eq:zor}
\text{where}\quad
Z_{\mathrm{or}}\,\lvert x \rangle 
&\;=\;
2\,\lvert 0^n \rangle \langle 0^n \rvert x\rangle \;-\; \lvert x \rangle
\;=\;
\begin{cases}
   -\,\lvert x \rangle & x \neq 0,\\
   \lvert x \rangle & \text{otherwise}.
\end{cases}
\end{align}
Above, $\langle x \rvert$ is the conjugate transpose of $\lvert x \rangle$.

Next, let us define two orthonormal vectors:
\begin{align}
\label{eq:average_s}
\lvert \psi_S \rangle 
&\;=\;
\frac{1}{\sqrt{n_s}}
\sum_{\substack{i=0 \\ i\in S}}^{2^n-1} \lvert i \rangle,
\\
\label{eq:average_ns}
\lvert \psi_{NS} \rangle
&\;=\;
\frac{1}{\sqrt{\,2^n - n_s\,}}
\sum_{\substack{i=0 \\ i\notin S}}^{2^n-1} \lvert i \rangle.
\end{align}
Using these, the initial state Eq.~\eqref{eq:uniform_superposition_state} can be expressed as a superposition of $\lvert \psi_S \rangle$ and $\lvert \psi_{NS} \rangle$:
\[
\lvert n \rangle 
\;=\;
\sqrt{\frac{n_s}{2^n}} \,\lvert \psi_S \rangle
\;+\;
\sqrt{\frac{2^n - n_s}{2^n}} \,\lvert \psi_{NS} \rangle.
\]

\begin{figure}[t]
  \centering
  \includegraphics[width=0.77\textwidth]{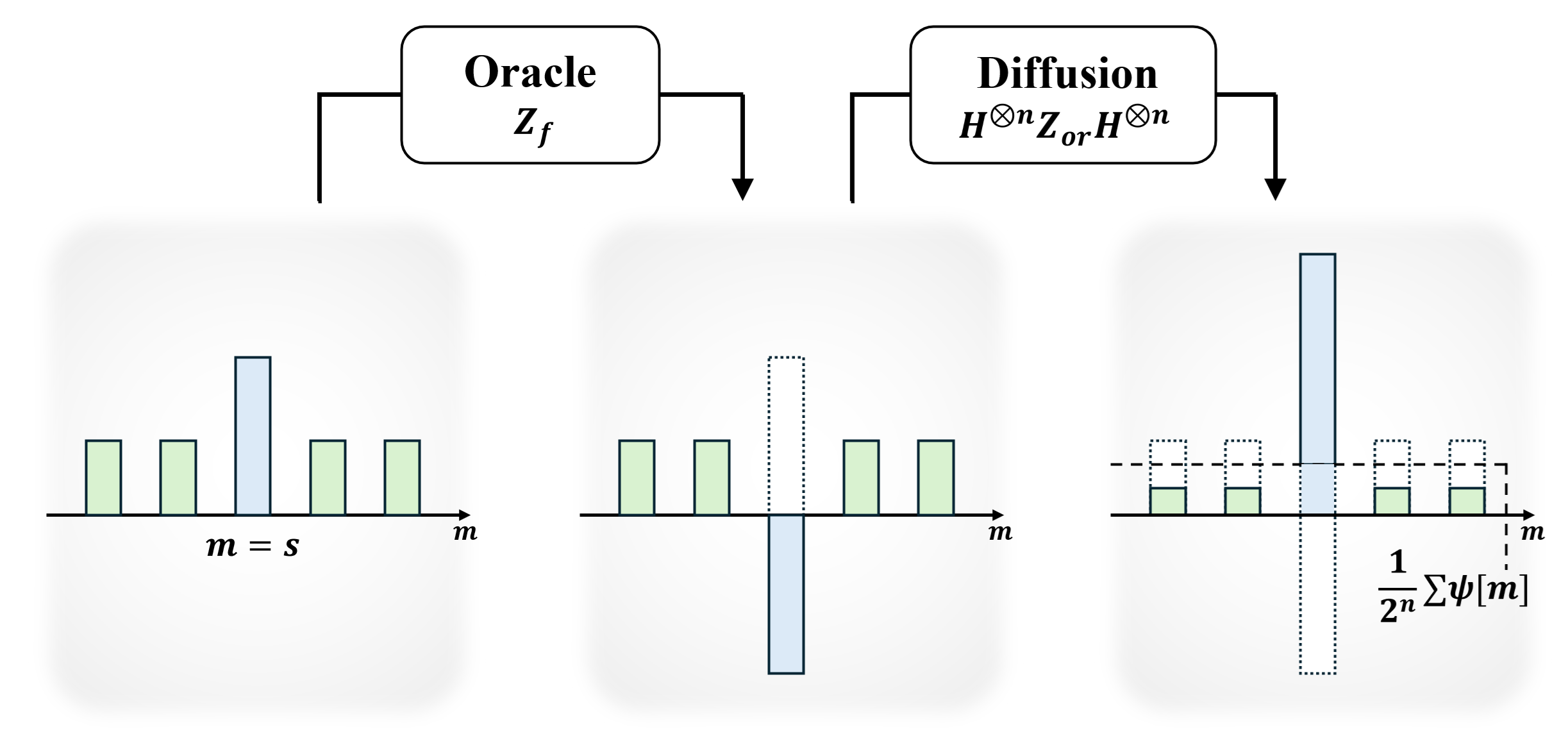}
  \caption{Illustration of the $G$ operation in Grover’s algorithm, which reflects the amplitudes of the state about their average to amplify the solution states.}
  \label{fig:G_op_graphic}
\end{figure}

From Eqs.~Eq.~\eqref{eq:average_s} and Eq.~\eqref{eq:average_ns}, the vectors $\lvert \psi_{S}\rangle$ and $\lvert \psi_{NS}\rangle$ form an orthonormal basis in a two-dimensional subspace, as they are mutually orthogonal and each is of unit length. Consequently, one can represent the initial state of Grover’s algorithm in this subspace, as shown in~\refFigure{fig:angular}, in \emph{angular form}:
\begin{align}
\lvert n\rangle
\;=\;
\sin(\theta)\,\lvert \psi_{S}\rangle
\;+\;
\cos(\theta)\,\lvert \psi_{NS}\rangle,
\quad
\text{where}
\quad
\theta
\;=\;
\sin^{-1}\!\Bigl(\!\sqrt{\tfrac{n_{s}}{2^{n}}}\Bigr).
\end{align}
Here, the Grover operator $G$ can be viewed as a rotation by $2\theta$ on the unit circle spanned by $\lvert \psi_{NS}\rangle$ and $\lvert \psi_{S}\rangle$. 
After $k$ applications of $G$, the state vector rotates by $2k\theta$, yielding a final state
\begin{align}
\label{eq:psi_final_angular}
\lvert \psi_{\mathrm{final}} \rangle
\;=\;
G^{k} \,\lvert n \rangle
\;=\;
\sin\bigl((2k+1)\,\theta\bigr)\,\lvert \psi_{S}\rangle
\;+\;
\cos\bigl((2k+1)\,\theta\bigr)\,\lvert \psi_{NS}\rangle.
\end{align}
Choosing $k$ according to Eq.~\eqref{eq:k} makes $(2k+1)\theta$ approach $\pi/2$, thereby maximizing the amplitude of $\lvert \psi_{S}\rangle$ and enabling a high-probability measurement of a solution. 
Letting $\delta = \tfrac{\pi}{2} - (2k+1)\theta$ denote the \emph{residual angle error}, one can show $|\delta| < \theta$, which implies
\begin{equation}\label{eq:residual}
	\left| \cos((2k+1)\theta) \right| = \left| \sin \delta \right| = \sin |\delta | < \sin \theta 
\end{equation}
Thus, the nonsolution amplitude remains strictly smaller than the initial amplitude $\sin(\theta)$, ensuring that the solution basis states are prominently amplified.

Because $\lvert \psi_S \rangle$ and $\lvert \psi_{NS} \rangle$ are orthonormal basis vectors in a two-dimensional subspace, the uniform state $\lvert n \rangle$ can be represented by an angular form $\sin \theta \,\lvert \psi_S \rangle + \cos \theta \,\lvert \psi_{NS} \rangle$, where $\theta$ satisfies $\sin^{-1} \theta = \sqrt{n_s / 2^n}$. Each application of $G$ effectively rotates the state by $2\theta$ within this plane. After $k$ iterations,
\begin{equation}
\label{eq:psi_final_angular}
\lvert \psi_{\mathrm{final}} \rangle
\;=\;
G^k \,\lvert n \rangle
\;=\;
\sin\bigl((2k+1)\,\theta\bigr)\,\lvert \psi_S \rangle
\;+\;
\cos\bigl((2k+1)\,\theta\bigr)\,\lvert \psi_{NS} \rangle,
\end{equation}
and, with $k$ chosen according to Eq.~\eqref{eq:k}, the amplitude for $\lvert \psi_S \rangle$ becomes near-maximal, yielding close to unity measurement probability for solution states.

\begin{figure}[t]
  \centering
  \includegraphics[width=0.5\textwidth]{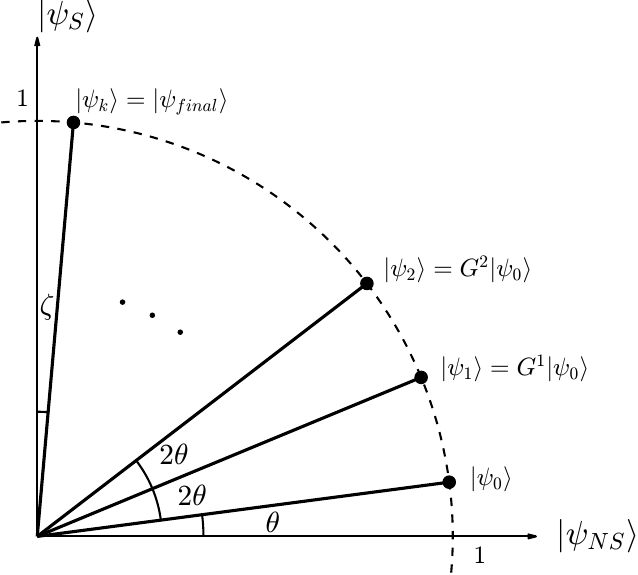}
  \caption{Geometric interpretation of $G$ as a rotation by $2\theta$ in the plane spanned by $\lvert \psi_{NS} \rangle$ and $\lvert \psi_S \rangle$.}
  \label{fig:angular}
\end{figure}

\subsection{Fixed-Point Arithmetic and Truncation Error}
\label{subsec:fixed_point}

\begin{figure}[b]
  \centering
  \includegraphics[width=0.8\textwidth]{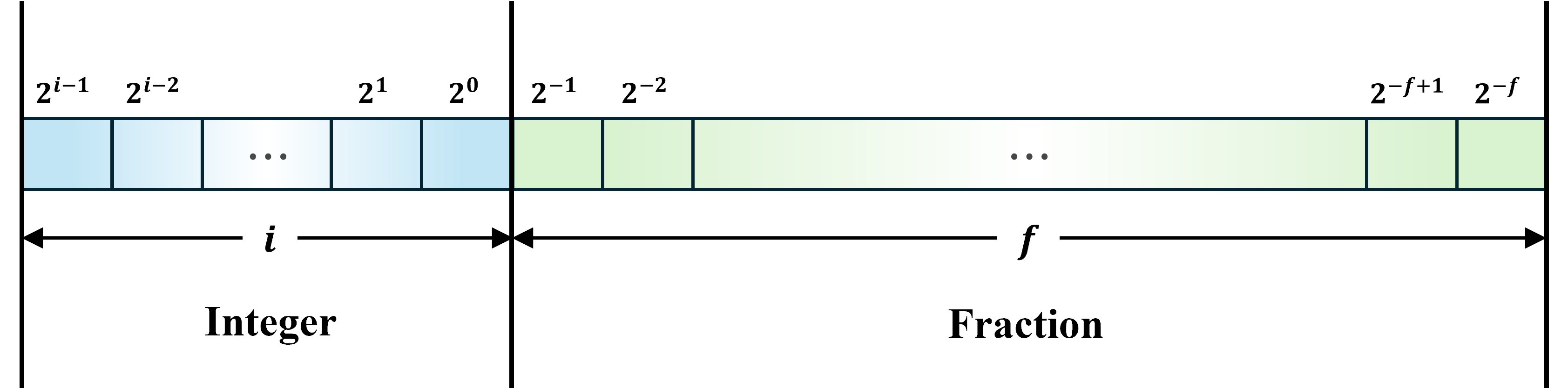}
  \caption{A real number in fixed-point format, where fixed integer and fractional bits define its range and precision.}
  \label{fig:fixed-point}
\end{figure}

\emph{Fixed-point arithmetic} encodes real numbers by assigning fixed bit widths for the integer and fractional parts, as illustrated in~\refFigure{fig:fixed-point}. 
This approach relies fundamentally on integer operations, often making it simpler and faster than floating-point arithmetic in hardware implementations. 
Denoting by $i$ and $f$ the bit widths for the integer and fractional parts, respectively, any real number $r$ in this system lies within
\begin{equation}
\label{eq:fp_range_and_bin}
-(2^{\,i-1} - 2^{-\,f})
\;\le\;
r
\;\le\;
2^{\,i-1} - 2^{-\,f},
\quad
\Delta r 
\;=\;
2^{-\,f}.
\end{equation}

\begin{figure}[t]
  \centering
  \includegraphics[width=0.82\textwidth]{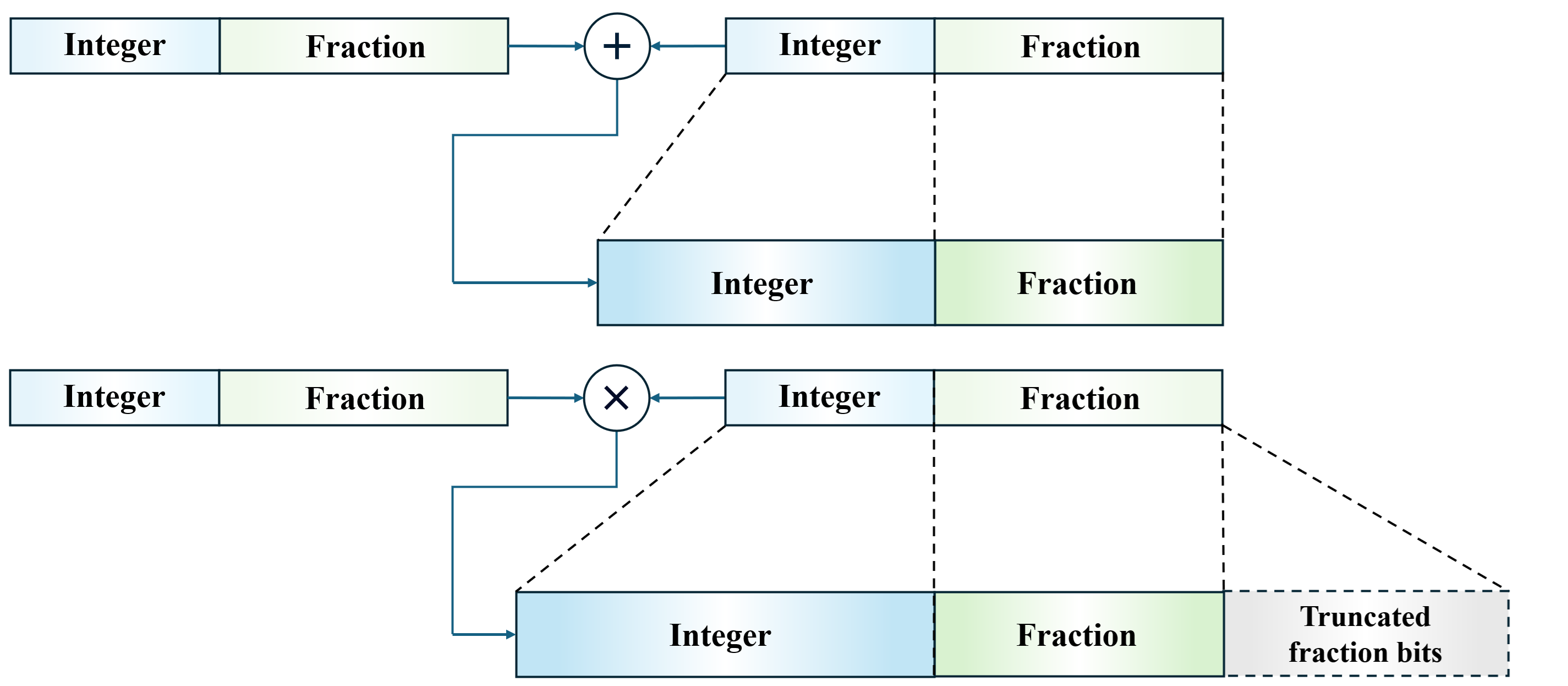}
  \caption{Changes in bit requirements during addition and multiplication in fixed-point arithmetic. Multiplying two fixed-point numbers may require twice as many fractional bits, causing truncation in the least significant bits.}
  \label{fig:fixed-point_arithmetics}
\end{figure}

\refFigure{fig:fixed-point_arithmetics} depicts how bit-space requirements differ for fixed-point addition and multiplication. 
For addition, if the integer portion exceeds the allotted bits, an overflow can occur, whereas the fractional precision remains unchanged. 
Consequently, no additional truncation error is introduced during addition. 
In multiplication, however, both the integer and fractional parts of the product can grow beyond the original bit widths. 
The integer part can overflow if it exceeds the allocated bits, and the fractional part can lose bits if its precision requirement (up to $2^{-2f}$) surpasses the available $2^{-f}$, resulting in \emph{truncation errors}.

In the specific case of Grover’s algorithm, one must handle an addition of $2^n$ amplitudes, each lying in $[-1,1]$, when performing the diffusion operation Eq.~\eqref{eq:diffusion_gate}. 
Hence, at least $n$ integer bits are necessary to prevent overflow in this summation. 
However, the algorithm’s multiplications do not exceed unity in magnitude, so no overflow arises there.

\begin{figure}[t]
  \centering
  \includegraphics[width=0.7\textwidth]{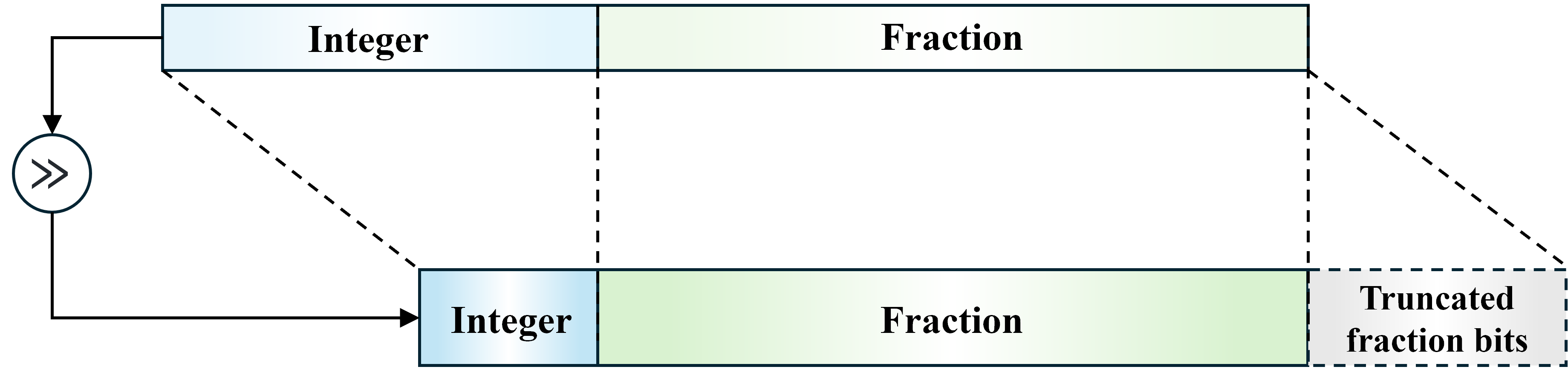}
  \caption{Right-shifting a fixed-point bit array. The truncated fractional bits become the source of truncation error.}
  \label{fig:right_bit-shift}
\end{figure}

Scaling operations (multiplying by powers of two) can often be handled by \emph{bit-shifting} instead of a more expensive multiplication logics (i.e. \emph{multipliers}, the high HW resource consumer). 
Indeed, left-shifting by $a$ bits corresponds to multiplying by $2^a$, while right-shifting by $a$ bits corresponds to multiplying by $2^{-a}$. 
As illustrated in~\refFigure{fig:right_bit-shift}, right-shifting can remove low-order fractional bits, thus generating truncation errors.

Mathematically, if a real $r$ in the arithmetic operation is represented as $r_{\textit{FP}}$ in fixed-point form with $f$ fractional bits, then the truncation error $\epsilon$ satisfies
\begin{equation}
\label{eq:r}
r \;=\; r_{\textit{FP}} + \epsilon,
\quad
\epsilon
\;=\;
r \;\bmod\; 2^{-\,f}.
\end{equation}
Hence, $r_{\textit{FP}}$ is a multiple of $2^{-\,f}$, while the residual $\epsilon$ captures the portion of $r$ that exceeds this precision limit. In the upcoming sections, we will analyze how such truncation errors accumulate when Grover’s algorithm is emulated under fixed-point arithmetic and evaluate their collective impact on the final measurement probability distribution.

\section{Theoretical Analysis of Truncation Errors in FP-Grover Emulation}
\label{sec:analysis}

\subsection{Preservation of Quantum State Structure under Grover’s Iterations}
\label{sec:proof}

From Eq.~\eqref{eq:grover_and_G} in Section~\ref{sec:prelim}, Grover’s operator $G$ selectively amplifies the amplitudes of the basis states corresponding to solutions. 
Moreover, from Eq.~\eqref{eq:uniform_superposition_state}, the initial state of Grover’s algorithm assigns identical amplitudes to all basis states. 
Consequently, at any step in Grover’s algorithm, the amplitudes of all solution states remain equal, and the amplitudes of all nonsolution states remain equal (albeit possibly at a different value). 
Formally, for an $n$-qubit system, let $\lvert \psi\rangle$ be the state of the system at any iteration. 
Then the amplitude of the $m^{th}$ basis state, $\psi[m]$, can be written as
\begin{equation}
\label{eq:psi_m}
\psi[m]
\;=\;
\begin{cases}
\psi_{S}, & m \in S,\\
\psi_{NS}, & \text{otherwise},
\end{cases}
\quad
m = 0,1,\ldots,2^n - 1,
\end{equation}
where $S$ is the set of solution indices.

In what follows, we leverage this simplified representation---the idea that any intermediate state has only two distinct amplitude values---to analyze truncation errors. We first prove, via mathematical induction, that Grover’s algorithm indeed preserves the form in~Eq.~\eqref{eq:psi_m} at every iteration.

\paragraph*{Base Case.}
Let $\lvert \psi_0\rangle = H^{\otimes n}\lvert 0^n\rangle$ be the initial state. From Eq.~\eqref{eq:uniform_superposition_state}, all basis states share the same amplitude
\begin{equation}
\label{eq:psi_0_m}
\psi_{0}[m]
\;=\;
\frac{1}{\sqrt{2^n}}, 
\quad
m = 0,1,\ldots,2^n - 1.
\end{equation}
Hence $\lvert \psi_0\rangle$ satisfies the structure of Eq.~\eqref{eq:psi_m}, with $\psi_S = \psi_{NS} = 1/\!\sqrt{2^n}$.

\paragraph*{Induction Step.}
Assume that, after $l$ iterations,
\begin{equation}
\label{eq:psi_l_m}
\psi_l[m]
\;=\;
\begin{cases}
\psi_{l,S}, & m \in S,\\
\psi_{l,NS}, & \text{otherwise}.
\end{cases}
\end{equation}
We must show the same holds for iteration $l+1$. By definition, 
\begin{equation}
\lvert \psi_{l+1}\rangle 
\;=\;
G\,\lvert \psi_l\rangle
\;=\;
\bigl(H^{\otimes n} Z_{\mathrm{or}} H^{\otimes n}\bigr)
\bigl(Z_f\bigr)\lvert \psi_l\rangle.
\end{equation}
First, let 
\(
\lvert \psi_l'\rangle = Z_f \,\lvert \psi_l\rangle
\).
Since $Z_f$ multiplies the amplitudes of solution states by $-1$ while leaving nonsolution states unchanged (see Eq.~\eqref{eq:oracle_gate}), the resulting state $\lvert \psi_l’\rangle$ again contains only two distinct amplitude values (with solution amplitudes negated).
Applying the diffusion operator $H^{\otimes n} Z_{\mathrm{or}} H^{\otimes n}$ defined in Eq.~\eqref{eq:diffusion_gate} to $\lvert \psi_l’\rangle$ computes the sum of all amplitudes, scales it by $2^{-n+1}$, and then subtracts each original amplitude individually, yielding:
\begin{align} 
	\notag
	 |\psi_{l+1}\rangle = G|\psi_{l}\rangle = (H^{\otimes n}Z_{or}H^{\otimes n})|\psi^{'}_{l}\rangle &= \left(\begin{bmatrix}
	2^{-n+1} & \hdots & 2^{-n+1} \\
	\vdots & \ddots & \vdots \\
	2^{-n+1} & \hdots & 2^{-n+1} 
	\end{bmatrix} - I_{n}\right)|\psi^{'}_{l}\rangle \\ \label{eq:psi_l+1}
	&= 2^{-n+1} \begin{bmatrix}
		\sum_{h=0}^{2^{n}-1} \psi^{'}_{l}[h] \\ \vdots \\ \sum_{h=0}^{2^{n}-1} \psi^{'}_{l}[h]
	\end{bmatrix} - |\psi^{'}_{l}\rangle
\end{align}
One can verify by direct inspection that this preserves the two-amplitude structure:
\begin{equation}
\label{eq:psi_l+1_m}
\psi_{l+1}[m]
\;=\;
2^{-\,n+1}
\sum_{h=0}^{2^n-1}
\psi_l'[h]
\;-\;
\psi_l'[m]
\;=\;
\begin{cases}
\displaystyle 2^{-\,n+1}\sum_{h=0}^{2^n-1} \psi_l'[h] + \psi_{l,S}, & m \in S,\\[6pt]
\displaystyle 2^{-\,n+1}\sum_{h=0}^{2^n-1} \psi_l'[h] - \psi_{l,NS}, & \text{otherwise}.
\end{cases}
\end{equation}
Thus, by mathematical induction, the state after each iteration $l$ continues to have exactly two distinct amplitude values, as per~Eq.~\eqref{eq:psi_m}.

\subsection{Numerical Formulation of Truncation Error Accumulation}
\label{subsec:numeric}

When using $f$ fractional bits in fixed-point arithmetic, any portion of a real number below $2^{-f}$ is truncated, which ultimately leads to distortions in the measured probabilities of a QC emulator. In this subsection, we develop a formal framework to track how these \emph{truncation errors} accumulate in FP-Grover emulation.

\paragraph*{Setup.}
Let 
$
\lvert \psi_l\rangle 
\;=\;
G^l\,\bigl(H^{\otimes n}\lvert 0^n\rangle\bigr)
$
denote the exact quantum state after $l$ applications of the Grover operator $G$. We represent each amplitude of $\lvert \psi_l\rangle$ in fixed-point form and gather them into a vector
$
\lvert \psi_{l,FP}\rangle.
$
Since truncation arises during each iteration of $G$ and propagates forward, let
$
\lvert \epsilon_l\rangle
$
capture the \emph{cumulative} truncation error up to the $l^{th}$ iteration. Formally,
\begin{equation}
\label{eq:psi_l_fp_error}
\lvert \psi_l\rangle
\;=\;
\lvert \psi_{l,FP}\rangle
\;+\;
\lvert \epsilon_l\rangle.
\end{equation}
From Section~\ref{sec:proof}, each Grover state takes the form in Eq.~\eqref{eq:psi_m}, so both 
\(\lvert \psi_{l,FP}\rangle\)
and 
\(\lvert \epsilon_l\rangle\)
retain two amplitude values according to whether \(m \in S\) or \(m \notin S\):
\begin{equation}
\label{eq:fp_error}
\psi_{l,FP}[m]
\;=\;
\begin{cases}
\psi_{l,FP,S}, & m \in S,\\
\psi_{l,FP,NS}, & \text{otherwise},
\end{cases}
\quad
\epsilon_l[m]
\;=\;
\begin{cases}
\epsilon_{l,S}, & m \in S,\\
\epsilon_{l,NS}, & \text{otherwise}.
\end{cases}
\end{equation}

\paragraph*{State Update.}
We now expand
\(\lvert \psi_{l+1}\rangle = G\,\lvert \psi_l\rangle\)
in fixed-point form:
\begin{align}
\label{eq:psi_l+1_fp_error}
\lvert \psi_{l+1}\rangle
&=\;
G\,\lvert \psi_{l,FP}\rangle
\;+\;
G\,\lvert \epsilon_l\rangle
\\[1pt]\notag
&=\;
\bigl(H^{\otimes n}Z_{\mathrm{or}}H^{\otimes n}\bigr)\,Z_f\,\lvert \psi_{l,FP}\rangle
\;+\;
\bigl(H^{\otimes n}Z_{\mathrm{or}}H^{\otimes n}\bigr)\,Z_f\,\lvert \epsilon_l\rangle
\\[1pt]\notag
&=\;
\bigl(H^{\otimes n}Z_{\mathrm{or}}H^{\otimes n}\bigr)\,\lvert \psi_{l,FP}'\rangle
\;+\;
\bigl(H^{\otimes n}Z_{\mathrm{or}}H^{\otimes n}\bigr)\,\lvert \epsilon_l'\rangle
\\[1pt]\notag
&=\;
\lvert \psi_{l,FP}''\rangle \;+\; \lvert \epsilon_l''\rangle,
\end{align}
where
\(\lvert \psi_{l,FP}'\rangle = Z_f\,\lvert \psi_{l,FP}\rangle\)
and
\(\lvert \epsilon_l'\rangle = Z_f\,\lvert \epsilon_l\rangle\).

\paragraph*{Detailed Recurrences.}
Consider first the fixed-point state vector
\(\lvert \psi_{l,FP}''\rangle = G\,\lvert \psi_{l,FP}\rangle\).
Applying the diffusion operator \((H^{\otimes n}Z_{\mathrm{or}}H^{\otimes n})\) involves summing all $2^n$ amplitudes, scaling by $2^{-n+1}$, and subtracting each original amplitude:
\begin{equation} \label{eq:psi_dprime_l_fp}
	|\psi^{''}_{l,FP}\rangle =  (H^{\otimes n}Z_{or}H^{\otimes n})|\psi^{'}_{l,FP}\rangle = 2^{-n+1}  \begin{bmatrix}
		\sum_{h=0}^{2^{n}-1} \psi^{'}_{l,FP}[h] \\ \vdots \\ \sum_{h=0}^{2^{n}-1} \psi^{'}_{l,FP}[h]
	\end{bmatrix} - |\psi^{'}_{l,FP}\rangle
\end{equation}
Addition and subtraction do not themselves degrade fractional precision in fixed-point arithmetic, but \emph{scaling} introduces a truncation error, denoted \(\epsilon_{l,scaled}'\).

Because \(\lvert \psi_{l,FP}'\rangle\) follows Eq.~\eqref{eq:fp_error}, we can factor out the number of solutions \(n_s\). For example,
\begin{align}
\label{eq:scale_sum_psi_prime_l_fp}
2^{-n+1}
\sum_{h=0}^{2^n-1} \psi_{l,FP}'[h]
&=\;
2^{-n+1}\Bigl((2^n - n_s)\,\psi_{l,FP,NS} - n_s\,\psi_{l,FP,S}\Bigr)
\\[3pt]\notag
&=\;
2\,\psi_{l,FP,NS}
\;-\;
2^{-n+1}\,n_s\bigl(\psi_{l,FP,NS} + \psi_{l,FP,S}\bigr),
\end{align}
which, in a practical fixed-point implementation, corresponds to a right bit-shift---specifically, the accumulated sum of the $2^n$ probability amplitudes is stored in a fixed-point number with an $n$-bit integer part, and the scaling operation is then performed by applying an $(n-1)$-bit right shift to this value---that may discard fractional bits below $2^{-f}$. We define
\begin{equation}
\label{eq:error_prime_l_scale}
\epsilon_{l,scaled}'
\;=\;
\Bigl(
2^{-n+1}\sum_{h=0}^{2^n-1} \psi_{l,FP}'[h]
\Bigr)
\;\bmod\;
2^{-f},
\end{equation}
reflecting the truncated remainder.

\paragraph*{Combined Fixed-Point Value and Truncation.}\,
Hence, the scaling result can be expressed as a combination of the fixed-point amplitude \(\psi_{l,FP,scaled}'\) and its truncation error \(\epsilon_{l,scaled}'\):
\begin{align}
\label{eq:scale_sum_psi_prime_l_fp_simple}
2^{-n+1}\sum_{h=0}^{2^n-1} \psi_{l,FP}'[h]
\;=\;
\psi_{l,FP,scaled}' + \epsilon_{l,scaled}'.
\end{align}
Combining Eq.~\eqref{eq:scale_sum_psi_prime_l_fp} and Eq.~\eqref{eq:scale_sum_psi_prime_l_fp_simple} yields
\begin{align}
\psi_{l,FP,scaled}'
&=\;
2^{-n+1}\sum_{h=0}^{2^n-1}\psi_{l,FP}'[h]
\;-\;
\epsilon_{l,scaled}' \notag
\\
&=\;
2\,\psi_{l,FP,NS}
\;-\;
2^{-n+1}\,n_s\,(\psi_{l,FP,NS} + \psi_{l,FP,S})
\;-\;
\epsilon_{l,scaled}'.
\end{align}
By substituting the above into Eq.~\eqref{eq:psi_dprime_l_fp} (along with Eq.~\eqref{eq:fp_error}), one obtains an explicit formula for
\(\psi_{l,FP}''[m]\), which can be written as
\begin{equation}
\label{eq:psi_dprime_l_fp_m}
\psi_{l,FP}''[m]
\;=\;
2^{-n+1}
\sum_{h=0}^{2^n-1}
\psi_{l,FP}'[h]
\;-\;
\psi_{l,FP}'[m]
\;=\;
\begin{cases}
\psi_{l,FP,scaled}' \;+\; \psi_{l,FP,S} \;+\;\epsilon_{l,scaled}', & m \in S,\\
\psi_{l,FP,scaled}' \;-\; \psi_{l,FP,NS} \;+\;\epsilon_{l,scaled}', & \text{otherwise}.
\end{cases}
\end{equation}

\paragraph*{Updated Error Terms.}
Eq.~\eqref{eq:psi_l+1_fp_error} similarly yields the updated error vector
\(\lvert \epsilon_l''\rangle = G\,\lvert \epsilon_l\rangle\). Namely,
\begin{equation}\label{eq:e_l_dprime}
	|\epsilon^{''}_{l}\rangle =  (H^{\otimes n}Z_{or}H^{\otimes n})|\epsilon^{'}_{l}\rangle = 2^{-n+1}  \begin{bmatrix}
		\sum_{h=0}^{2^{n}-1} \epsilon^{'}_{l}[h] \\ \vdots \\ \sum_{h=0}^{2^{n}-1} \epsilon^{'}_{l}[h]
	\end{bmatrix} - |\epsilon^{'}_{l}\rangle
\end{equation}
where
\begin{align*}
2^{-n+1}
\sum_{h=0}^{2^n-1}
\epsilon_l'[h]
&=\;
2^{-n+1}
\Bigl(
(2^n - n_s)\,\epsilon_{l,NS}
\;-\;
n_s\,\epsilon_{l,S}
\Bigr)
\\[3pt]
&=\;
2\,\epsilon_{l,NS}
\;-\;
2^{-n+1}\,n_s\,(\epsilon_{l,NS} + \epsilon_{l,S}).
\end{align*}
Adding the scaling-induced error \(\epsilon_{l,scaled}'\) finally leads to the recurrences for the next iteration $l+1$. In particular, we have
\begin{align}\label{eq:psi_fp_l+1}
	\psi_{l+1,FP}[m] = \psi^{''}_{l,FP}[m] -  \epsilon^{'}_{l,scaled} 
	= \begin{cases}  
		\psi^{'}_{l,FP,scaled} + \psi_{l,FP,S} & \text{if} \ m \in S \\ 
		\psi^{'}_{l,FP,scaled} - \psi_{l,FP,NS} & \text{otherwise,}
	\end{cases}
\end{align}
\begin{align}\label{eq:error_l+1}
	\notag
	\epsilon_{l+1}[m] &= \epsilon^{''}_{l}[m] + \epsilon^{'}_{l,scaled} \\
	&= \begin{cases}  
		\{2\epsilon_{l,NS} - 2^{-n+1}n_{s}(\epsilon_{l,NS} + \epsilon_{l,S}) + \epsilon_{l,S}\} + \epsilon^{'}_{l,scaled} & \text{if} \ m \in S \\
		\{2\epsilon_{l,NS} - 2^{-n+1}n_{s}(\epsilon_{l,NS} + \epsilon_{l,S}) - \epsilon_{l,NS}\} + \epsilon^{'}_{l,scaled} & \text{otherwise,}
	\end{cases}
\end{align}
with initial values
\begin{equation}
\label{eq:psi_e_initial}
\psi_{0,FP}[m]
\;=\;
\frac{1}{\sqrt{2^n}} \;-\;\epsilon_0[m],
\quad
\epsilon_0[m]
\;=\;
\Bigl(\tfrac{1}{\sqrt{2^n}}\Bigr)\bmod 2^{-f}.
\end{equation}
Iterating Eq.~\eqref{eq:psi_fp_l+1} and Eq.~\eqref{eq:error_l+1} through $l = k$ yields the final state 
\(\lvert \psi_{k,FP}\rangle\)
and the accumulated error
\(\lvert \epsilon_k\rangle\)
in the FP-Grover emulation.

\paragraph*{Approximate Scaling Behavior.}
To see how $\lvert \epsilon_k\rangle$ scales with $n$ and $f$, consider sufficiently large $n$ and small $n_s$, so $2^{-n+1}n_s \approx 0$. Under this approximation, the update Eq.~\eqref{eq:error_l+1} reduces to
\[
\epsilon_{l+1}[m]
\;=\;
\begin{cases}
2\,\epsilon_{l,NS}
\;+\;
\epsilon_{l,S}
\;+\;
\epsilon_{l,scaled}', & m \in S,\\
\epsilon_{l,NS}
\;+\;
\epsilon_{l,scaled}', & \text{otherwise}.
\end{cases}
\]
Hence,
\begin{align}
\label{eq:error_k_ns_bigo}
\epsilon_{k,NS}
&=\;
\epsilon_{k-1,NS} \;+\; \epsilon_{k-1,scaled}'
\;=\;
\epsilon_{k-2,NS} \;+\; \epsilon_{k-2,scaled}' \;+\; \epsilon_{k-1,scaled}'
\;=\;
\hdots
\;=\;
\epsilon_{0,NS}
\;+\;
\sum_{h=0}^{k-1}
\epsilon_{h,scaled}'.
\end{align}
Since $\epsilon_{0,NS}<2^{-f}$ and each $\epsilon_{h,scaled}'<2^{-f}$ (from Eq.~\eqref{eq:psi_e_initial} and Eq.~\eqref{eq:error_prime_l_scale}), we have
\begin{align}\label{eq:epsilon_k_NS_bigO}
	\epsilon_{k,NS} = O(2^{-f}) + O(k\cdot2^{-f}) = O(2^{\frac{n}{2}-f}),  
\end{align}
where $k \approx O(2^{\tfrac{n}{2}})$ by Eq.~\eqref{eq:k}. Meanwhile, the solution error $\epsilon_{k,S}$ grows even more aggressively:
\begin{align}
\label{eq:epsilon_k_S_bigO}
\epsilon_{k,S}
&=\;
\epsilon_{k-1,S} 
+ 
\bigl(2\,\epsilon_{k-1,NS} + \epsilon_{k-1,scaled}'\bigr)
\;=\;
\hdots
\;=\;
\epsilon_{0,S}
\;+\;
\sum_{m=0}^{k-1}
\bigl(2\,\epsilon_{m,NS} + \epsilon_{m,scaled}'\bigr)
\\[-2pt]\notag
&=\;
O\bigl(k^2\,2^{-f}\bigr)
\;=\;
O\bigl(2^{\,n-f}\bigr).
\end{align}
These results will be central to our analysis in Section~\ref{subsec:measurement_error} on the total measurement error in the final FP-Grover state.

\subsection{Impact of Truncation Errors on Measurement Probability Distribution}
\label{subsec:measurement_error}

We now use the big-O bounds derived in Section~\ref{subsec:numeric} for the errors in the quantum-state coefficients, \(\epsilon_{k,NS}\) and \(\epsilon_{k,S}\) (Eqs.~\eqref{eq:epsilon_k_NS_bigO}--\eqref{eq:epsilon_k_S_bigO}), to determine how these errors ultimately distort the \emph{measurement probability distribution} produced by the FP-Grover emulation. 
Conceptually, the severity of an emulation error corresponds to the discrepancy between its output probability distribution and the ideal distribution. 
Hence, we adopt the $\ell_2$ distance between the true probabilities and the emulated probabilities as a representative metric, which we refer to as the \emph{$\ell_2$ error}.

\paragraph*{Definition of $\ell_2$ Error.}
Let $p[m]$ be the ideal probability of measuring basis state $m$ in the final state after $k$ Grover iterations, and let $p_{\textit{FP}}[m]$ be the corresponding probability under fixed-point emulation. We define the $\ell_2$ error as
\begin{equation}
\label{eq:L2_experiment}
\ell_{2}
\;=\;
\sqrt{\sum_{m=0}^{2^n-1} \bigl(p[m] - p_{\textit{FP}}[m]\bigr)^{2}}.
\end{equation}
Here, $p[m]$ follows from Eq.~\eqref{eq:prob_is_square}, which states
\begin{equation}
p[m] = \bigl\lvert \psi_{k}[m]\bigr\rvert^{2} = \bigl(\psi_{k,FP}[m] + \epsilon_{k}[m]\bigr)^{2} = \bigl(\psi_{k,FP}[m]\bigr)^{2} + 2\,\psi_{k,FP}[m]\,\epsilon_{k}[m] + \bigl(\epsilon_{k}[m]\bigr)^{2}.
\label{eq:p_m}
\end{equation}
Although $\bigl(\psi_{k,FP}[m]\bigr)^2$ itself may exceed the fixed-point precision $2^{-f}$, we capture this mismatch by a \emph{square} truncation error $\epsilon_{sq}[m]$:
\begin{align}
\label{eq:p_fp_m}
p_{\textit{FP}}[m]
&=\;
\bigl(\psi_{k,FP}[m]\bigr)^2
\;-\;
\epsilon_{sq}[m],
\\
\label{eq:e_sq_m}
\epsilon_{sq}[m]
&=\;
\Bigl(\bigl(\psi_{k,FP}[m]\bigr)^2\Bigr)
\;\bmod\;
2^{-f}.
\end{align}
Then,
\[
p[m]
\;=\;
p_{\textit{FP}}[m]
\;+\;
\epsilon_{sq}[m]
\;+\;
2\,\psi_{k,FP}[m]\,\epsilon_{k}[m]
\;+\;
\bigl(\epsilon_{k}[m]\bigr)^2,
\]
so the total truncation error in $p[m]$ is
\begin{equation}
\label{eq:e_p_m}
\epsilon_{p}[m]
\;=\;
\epsilon_{sq}[m]
\;+\;
2\,\psi_{k,FP}[m]\,\epsilon_{k}[m]
\;+\;
\bigl(\epsilon_{k}[m]\bigr)^2.
\end{equation}

\paragraph*{Scaling Behavior of Probability Errors.}
From Eq.~\eqref{eq:e_sq_m}, we know $\epsilon_{sq}[m] = O\bigl(2^{-f}\bigr)$. 
Moreover, if $m \notin S$, we have $\lvert \psi_{k,FP}[m]\rvert = \lvert \psi_{k,FP,NS}\rvert$, which can be expressed as:
\begin{equation} \label{eq:abs_psi_k_FP_NS}
|\psi_{k,FP,NS}| \approx |\psi_{k,NS}| 
= \frac{\left|\cos((2k+1)\theta)\right|}{\sqrt{2^{n}-n_{s}}}
< \frac{\sin\theta}{\sqrt{2^{n}-n_{s}}}
= \frac{\sqrt{n_{s}}}{\sqrt{2^{n}}\sqrt{2^{n}-n_{s}}},
\quad \text{where}\quad \sin^{-1}\theta = \sqrt{\frac{n_{s}}{2^{n}}}.
\end{equation}
Thus, we conclude $|\psi_{k,FP,NS}| = O(2^{-n})$.
Conversely, for $m \in S$, $\lvert \psi_{k,FP}[m]\rvert = \lvert \psi_{k,FP,S}\rvert \approx n_s^{-1}$, effectively a constant $O(1)$ with respect to $n$ and $f$. 
Including the bounds for $\epsilon_{k,NS}$ and $\epsilon_{k,S}$ derived in Section~\ref{subsec:numeric}, we obtain:
\begin{align*}
\epsilon_{sq,NS} &= O\bigl(2^{-f}\bigr),
&&
\epsilon_{sq,S} = O\bigl(2^{-f}\bigr), \\
\lvert \psi_{k,FP,NS}\rvert &= O\bigl(2^{-n}\bigr),
&&
\lvert \psi_{k,FP,S}\rvert = O(1), \\
\epsilon_{k,NS} &= O\bigl(2^{\tfrac{n}{2}-f}\bigr),
&&
\epsilon_{k,S} = O\bigl(2^{\,n-f}\bigr).
\end{align*}
Substituting these into Eq.~\eqref{eq:e_p_m} isolates the scaling behavior of the probability error \(\epsilon_{p}[m]\). For nonsolution states:
\begin{align}
\label{eq:e_p_NS_bigO_mid}
\epsilon_{p,NS}
&=\;
O\bigl(2^{-f}\bigr)
\;\pm\;
O\bigl(2^{-n}\bigr)\,O\bigl(2^{\tfrac{n}{2}-f}\bigr)
\;+\;
O\bigl(2^{2(\tfrac{n}{2}-f)}) 
\\\notag
&=\;
O\bigl(2^{-f}\bigr)
\;\pm\;
O\bigl(2^{-\tfrac{n}{2}-f}\bigr)
\;+\;
O\bigl(2^{\,n-2f}\bigr).
\end{align}
If $n<f$, then $O(2^{-f})$ dominates, so
\begin{equation}
\label{eq:e_p_NS_bigO}
\epsilon_{p,NS}
\;=\;
O\bigl(2^{-f}\bigr).
\end{equation}
Likewise, for solution states,
\begin{align}
\label{eq:e_p_S_bigO}
\epsilon_{p,S}
&=\;
O\bigl(2^{-f}\bigr)
\;+\;
O(1)\,O\bigl(2^{\,n-f}\bigr)
\;+\;
O\bigl(2^{2(n-f)}\bigr)
\;=\;
O\bigl(2^{\,n-f}\bigr).
\end{align}

\paragraph*{$\ell_2$ error of the Final Distribution.}
Because $\epsilon_{p}[m]$ follows the same structure as in Eq.~\eqref{eq:psi_m}, we can substitute it directly into Eq.~\eqref{eq:L2_experiment}:
\begin{equation}
\label{eq:L2_theory}
\ell_{2}
\;=\;
\sqrt{\,
\sum_{m=0}^{\,2^n-1}
\bigl(\epsilon_{p}[m]\bigr)^2
}
\;=\;
\sqrt{\,
(2^n - n_s)\,\epsilon_{p,NS}^2
\;+\;
n_s\,\epsilon_{p,S}^2
}.
\end{equation}
Assuming $n_s \ll 2^n$, we apply the bounds from Eqs.~\eqref{eq:e_p_NS_bigO}--\eqref{eq:e_p_S_bigO} to obtain
\begin{equation}
\label{eq:L2_big_O}
\ell_{2}
\;=\;
\sqrt{\,
O\bigl(2^n\bigr)\,O\bigl(2^{-2f}\bigr)
\;+\;
O\bigl(2^{2(n-f)}\bigr)
}
\;\approx\;
\sqrt{\,
O\bigl(2^{2(n-f)}\bigr)
}
\;=\;
O\bigl(2^{\,n-f}\bigr).
\end{equation}
Hence, each additional fractional bit (increasing $f$ by 1) effectively halves $\ell_{2}$, whereas each extra qubit (increasing $n$ by 1) doubles it, provided $n_s$ remains relatively small compared to $2^n$. This completes our theoretical analysis, giving a clear guideline on how $n$ and $f$ jointly drive the magnitude of truncation-induced measurement errors in FP-Grover emulation.

\section{Experimental Validation and Precision Optimization for FP-Grover Emulation}
\label{sec:exp}

\subsection{Verification of the Proposed Truncation Error Formulation}
\label{subsec:validation}

\begin{algorithm}[b]
\small
\caption{Arithmetic procedures for emulating Grover’s algorithm using fixed-point and double-precision floating-point expressions, enabling direct evaluation of the $\ell_2$ truncation error.}\label{alg:experiment}
\begin{tabular}{p{0.45\textwidth}|p{0.49\textwidth}}
\begin{algorithmic}[1]
\STATE \textbf{procedure} GROVER\_FIXED$(n, f, n_{s})$
\STATE \quad \textbf{int} $amp_{fixed,NS}, amp_{fixed,S} = \left\lfloor2^{-n/2 + f}\right\rfloor$; \hspace{5pt}// fixed-point expression of probability amplitude
\STATE \quad \textbf{int} $sum$;
\STATE \quad \textbf{int} $k =\left\lfloor \frac{\pi}{4}\sqrt{\frac{2^{n}}{n_{s}}}\right\rceil$; \hspace{5pt}// number of iteration of $G$
\STATE \quad \textbf{for} $i = 0$ to $k-1$ \textbf{do}
\STATE \quad \quad $amp_{fixed,S} = -amp_{fixed,S}$;
\STATE \quad \quad $sum = (2^{n} - n_{s}) * amp_{fixed,NS} + n_{s} * amp_{fixed,S}$;
\STATE \quad \quad $amp_{fixed,NS} = sum >> (n+1)- amp_{fixed,NS}$;
\STATE \quad \quad $amp_{fixed,S} = sum >> (n+1) - amp_{fixed,S}$;
\STATE \quad \textbf{end for} 
\STATE \quad \textbf{return} $amp_{fixed,NS}, amp_{fixed,S}$;
\STATE \textbf{end procedure}
\end{algorithmic}
&
\begin{algorithmic}[1]
\STATE \textbf{procedure} GROVER\_DOUBLE$(n, n_{s})$
\STATE \quad \textbf{double} $amp_{NS}, amp_{S} = 2^{-n/2}$; \hspace{5pt} // double-precision floating-point expression of probability amplitude
\STATE \quad \textbf{double} $sum$;
\STATE \quad \textbf{int} $k =\left\lfloor \frac{\pi}{4}\sqrt{\frac{2^{n}}{n_{s}}}\right\rceil$;\hspace{5pt}// number of iteration of $G$
\STATE \quad \textbf{for} $i = 0$ to $k-1$ \textbf{do}
\STATE \quad \quad $amp_{S} = -amp_{S}$;
\STATE \quad \quad $sum = (2^{n} - n_{s}) * amp_{NS} + n_{s} * amp_{S}$;
\STATE \quad \quad $amp_{NS} = 2^{-n+1} * sum - amp_{NS}$;
\STATE \quad \quad $amp_{S} = 2^{-n+1} * sum - amp_{S}$;
\STATE \quad \textbf{end for}
\STATE \quad \textbf{return} $amp_{NS}, amp_{S}$;
\STATE \textbf{end procedure}
\end{algorithmic}

\\ \hline
\multicolumn{2}{p{0.90\textwidth}}{
\begin{algorithmic}[1]
\STATE \textbf{procedure} GROVER\_L2$(n, f, n_{s})$
\STATE \quad \textbf{int} $amp_{fixed,NS}, amp_{fixed_S} = GROVER\_FIXED(n, f, n_{s})$;
\STATE \quad \textbf{double} $amp_{NS}, amp_{S} = GROVER\_DOUBLE(n, n_{s})$;
\STATE \quad \textbf{double} $prob_{fixed,NS}, prob_{fixed,S}, prob_{NS}, prob_{S}$; \hspace{5pt} // measurement probability
\STATE \quad $prob_{fixed,NS} = (\textbf{double})(amp_{fixed,NS}^2 >> f) * 2^{-f}$;
\STATE \quad $prob_{fixed,S} = (\textbf{double})(amp_{fixed,S}^2 >> f) * 2^{-f}$;
\STATE \quad $prob_{NS} = amp_{NS}^2,\hspace{5pt}prob_{S} = amp_{S}^2$;
\STATE \quad \textbf{return} $L_{2} = \sqrt{(2^{n}-n_{s})(prob_{NS}-prob_{fixed,NS})^{2} + n_{s}(prob_{S}-prob_{fixed,S})^{2}}$;
\STATE \textbf{end procedure}
\end{algorithmic}
}
\end{tabular}
\end{algorithm}

\begin{algorithm}
\small
\caption{Arithmetic procedure that implements the proposed theoretical framework for computing the $\ell_2$ error in FP-Grover emulation.}\label{alg:proposed}
\begin{tabular}{p{0.9\textwidth}}
\begin{algorithmic}[1]
\STATE \textbf{procedure} L2\_THEORETICAL$(n, f, n_{s})$
\STATE \quad \textbf{double} $\psi_{k,FP,NS}, \psi_{k,FP,S} = 2^{-n/2} - \mathrm{mod}(2^{-n/2}, 2^{-f})$;\hspace{5pt}// probability amplitude in fixed-point expression
\STATE \quad \textbf{double} $\epsilon_{k,NS}, \epsilon_{k,S} = \mathrm{mod}(2^{-n/2}, 2^{-f})$;\hspace{5pt}// accumulated truncation error
\STATE \quad \textbf{double} $\psi_{scaled}$, $\epsilon_{scaled}$
\STATE \quad \textbf{int} $k =\left\lfloor \frac{\pi}{4}\sqrt{\frac{2^{n}}{n_{s}}}\right\rceil$;\hspace{5pt}// number of iteration of $G$
\STATE \quad \textbf{for} $i = 0$ to $k-1$ \textbf{do}
\STATE \quad \quad $\psi_{scaled} = 2*\psi_{FP,NS} - 2^{-n+1}*n_{s}*(\psi_{FP,NS}+\psi_{FP,NS})$;
\STATE \quad \quad $\epsilon_{scaled} = \mathrm{mod}(\psi_{scaled}, 2^{-f})$;
\STATE \quad \quad $\psi_{scaled} = \psi_{scaled} - \epsilon_{scaled}$;
\STATE \quad \quad $\psi_{k,FP,NS} = \psi_{scaled} - \psi_{k,FP,NS}$;
\STATE \quad \quad $\psi_{k,FP,S} = \psi_{scaled} - \psi_{k,FP,S}$;
\STATE \quad \quad $\epsilon_{k,NS} = \{2\epsilon_{k,NS} - 2^{-n+1}n_{s}(\epsilon_{k,NS} + \epsilon_{k,S}) - \epsilon_{k,NS}\} + \epsilon_{scaled}$;
\STATE \quad \quad $\epsilon_{k,S} = \{2\epsilon_{k,S} - 2^{-n+1}n_{s}(\epsilon_{k,NS} + \epsilon_{k,S}) + \epsilon_{k,S}\} + \epsilon_{scaled}$; 
\STATE \quad \textbf{end for}
\STATE \quad $\epsilon_{sq,NS} = \mathrm{mod}(\psi_{k,FP,NS}^{2}, 2^{-f})$;
\STATE \quad $\epsilon_{sq,S} = \mathrm{mod}(\psi_{k,FP,S}^{2}, 2^{-f})$;
\STATE \quad $\epsilon_{p,NS} = \epsilon_{sq,NS} + 2*\psi_{k,FP,NS}*\epsilon_{k,NS} + \epsilon_{k,NS}^{2}$;
\STATE \quad $\epsilon_{p,S} = \epsilon_{sq,S} + 2*\psi_{k,FP,S}*\epsilon_{k,S} + \epsilon_{k,S}^{2}$;
\STATE \quad \textbf{return} $L_{2} = \sqrt{(2^{n}-n_{s})\epsilon_{p,NS}^{2} + n_{s}\epsilon_{p,S}^{2}}$;
\STATE \textbf{end procedure}
\end{algorithmic}
\end{tabular}
\end{algorithm}

To confirm that the theoretical $\ell_2$ error analysis derived in previous sections accurately reflects the actual behavior of an FP-Grover emulation, we must verify that the \emph{theoretical} errors match the \emph{measured} errors in a practical implementation. 
To this end, we designed \refAlgo{alg:experiment}, which computes the final amplitudes of Grover’s algorithm and evaluates the resulting $\ell_2$ error via both fixed-point and double-precision floating-point expressions.

\paragraph*{Algorithm~\ref{alg:experiment}.}
The procedure \textsc{GROVER\_FIXED} computes the final quantum-state amplitudes $(amp_{fixed,NS},\,amp_{fixed,S})$ in a fixed-point representation, while \textsc{GROVER\_DOUBLE} obtains the exact double-precision floating-point amplitudes $(amp_{NS},\,amp_{S})$. 
Both subroutines implement $k$ iterations of Grover’s operator $G$ (see Eq.~\eqref{eq:grover_and_G}). 
After computing these amplitudes, the procedure \textsc{GROVER\_L2} calculates the \emph{measured} $\ell_2$ error by comparing the resulting probabilities to the ground-truth probabilities:
\begin{equation*}
\ell_2 
\;=\;
\sqrt{\,
(2^{n} - n_{s})
\bigl(\,prob_{NS} - prob_{fixed,NS}\bigr)^{2}
\;+\;
n_{s}\,\bigl(\,prob_{S} - prob_{fixed,S}\bigr)^{2}
},
\end{equation*}
corresponding to Eq.~\eqref{eq:L2_experiment}.

\paragraph*{Algorithm~\ref{alg:proposed}.}
We also implement \textsc{L2\_THEORETICAL}, which computes the \emph{theoretical} $\ell_2$ error based on the formulas derived in Sections~\ref{subsec:numeric} and \ref{subsec:measurement_error}. Concretely, this algorithm:

\begin{enumerate}
\item Uses Eqs.~\eqref{eq:psi_fp_l+1} and \eqref{eq:error_l+1} to iteratively compute the final fixed-point amplitudes $(\psi_{k,FP,NS},\,\psi_{k,FP,S})$ and the accumulated truncation errors $(\epsilon_{k,NS},\,\epsilon_{k,S})$.  
\item Evaluates the measurement probability error $(\epsilon_{p,NS},\,\epsilon_{p,S})$ via Eq.~\eqref{eq:e_p_m}.  
\item Substitutes these results into Eq.~\eqref{eq:L2_theory} to produce the theoretical $\ell_2$ error value.
\end{enumerate}

\begin{figure}[t]
	\centering
	\includegraphics[width=0.78\textwidth]{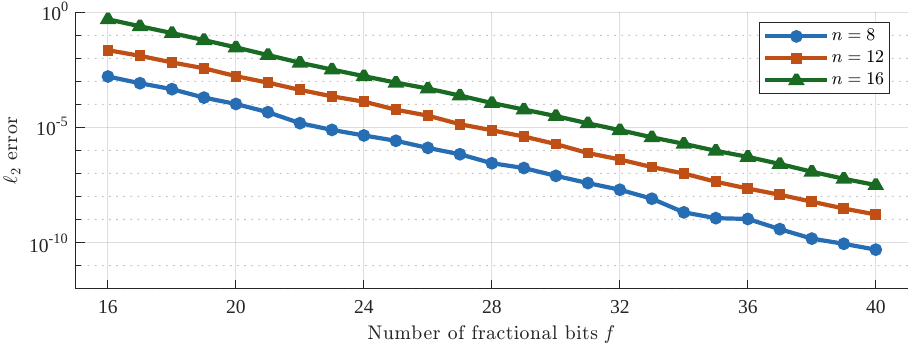}
	\caption{$\ell_2$ error for varying numbers of fractional bits $f$, evaluated at different qubit counts ($n=8,12,16$). Note that the vertical axis is in logarithmic scale.}
	\label{fig:l2_graph}
\end{figure}

\paragraph*{Experimental Results.}
We tested Algorithm \ref{alg:experiment} and \ref{alg:proposed} by varying $n=8, 12, 16$, $f=16, 17,\dots,40$, and setting $n_s=1$. 
For each parameter set $(n,f,n_s)$, we compared the \emph{measured} error (from the actual fixed-point emulator) to the \emph{theoretical} error (from our derived formulas). 
As shown in \refFigure{fig:l2_graph}, which plots $\ell_2$ error versus fractional bit count $f$ on a semi-log scale, these two values agree precisely across all tested parameters, confirming that our error formulation in Eq.~\eqref{eq:L2_theory} accurately characterizes truncation errors in FP-Grover emulation.

\begin{table}[t]
\begin{center}
\caption{$\ell_2$ error for selected $f$ with the range width $w=4$, plus the average ratios $r_{prec}(f)$ and $r_{num\_q}$ within each range.}
\small
\renewcommand{\arraystretch}{1.5}
	\centering
	\begin{tabular}{c|c|ccccccc}
\toprule
\multicolumn{2}{c|}{\multirow{2}{*}{$n$}} & \multicolumn{7}{c}{$f$} \\
\cline{3-9}
\multicolumn{2}{c|}{~} & 16 & 20 & 24 & 28 & 32 & 36 & 40 \\
\midrule
\multirow{2}{*}{8} & $\ell_{2}$ & $1.618 \times 10^{-3}$ & $1.039 \times 10^{-4}$ & $4.459 \times 10^{-6}$ & $2.781 \times 10^{-7}$ & $1.959 \times 10^{-8}$ & $1.057 \times 10^{-9}$ & $4.900 \times 10^{-11}$ \\
& $r_{prec}(f)$ & \cellcolor{orange!10} $-$ & \cellcolor{orange!10} $0.5034$ & \cellcolor{orange!10} $0.4551$ & \cellcolor{orange!10} $0.4997$ & \cellcolor{orange!10} $0.5152$ & \cellcolor{orange!10} $0.4820$ & \cellcolor{orange!10} $0.4640$ \\
\hline
\multirow{2}{*}{12} & $\ell_{2}$ & $2.250 \times 10^{-2}$ & $1.677 \times 10^{-3}$ & $1.311 \times 10^{-4}$ & $7.490 \times 10^{-6}$ & $4.048 \times 10^{-7}$ & $2.212 \times 10^{-8}$ & $1.643 \times 10^{-9}$ \\
& $r_{prec}(f)$ & \cellcolor{orange!10} $-$ & \cellcolor{orange!10} $0.5225$ & \cellcolor{orange!10} $0.5287$ & \cellcolor{orange!10} $0.4890$ & \cellcolor{orange!10} $0.4821$ & \cellcolor{orange!10} $0.4835$ & \cellcolor{orange!10} $0.5221$ \\
\hline
\multirow{2}{*}{16} & $\ell_{2}$ & $4.936 \times 10^{-1}$ & $2.980 \times 10^{-2}$ & $1.675 \times 10^{-3}$ & $1.143 \times 10^{-4}$ & $7.599 \times 10^{-6}$ & $5.243 \times 10^{-7}$ & $3.121 \times 10^{-8}$ \\
& $r_{prec}(f)$ & \cellcolor{orange!10} $-$ & \cellcolor{orange!10} $0.4957$ & \cellcolor{orange!10} $0.4869$ & \cellcolor{orange!10} $0.5112$ & \cellcolor{orange!10} $0.5078$ & \cellcolor{orange!10} $0.5125$ & \cellcolor{orange!10} $0.4940$ \\
\midrule
\multicolumn{2}{c|}{$r_{num\_q}(f)$} & \cellcolor{orange!10} $2.0442$ & \cellcolor{orange!10} $2.0285$ & \cellcolor{orange!10} $2.0981$ & \cellcolor{orange!10} $2.1220$ & \cellcolor{orange!10} $2.1067$ & \cellcolor{orange!10} $2.1724$ & \cellcolor{orange!10} $2.2415$ \\
\bottomrule
	\end{tabular}
	\label{table:l2_ratio}
\end{center}
\end{table}

Notably, \refFigure{fig:l2_graph} shows linear decreases of the error curves for each $n$. 
These linear trends confirm the \emph{exponential} falloff predicted by the big-O relationship derived in Eq.~\eqref{eq:L2_big_O}. 
To numerically verify these observations, \refTable{table:l2_ratio} partitions the $\ell_2$ error data into intervals of width $w_{prec} = 4$ in terms of $f$ and computes the average \emph{scaling ratio} within each interval. 
For an interval ending at fractional bit $f = a$, the average scaling ratio $r_{prec}(a)$ is defined as
\begin{equation} 
	r_{prec}(a) = \left(\frac{L_{2}\big|_{f=a}}{L_{2}\big|_{f=a-w_{prec}}}\right)^{\frac{1}{w_{prec}}},
\end{equation}
which quantitatively characterizes how quickly the $\ell_2$ error decreases as $f$ increases. 
Across all intervals tested, the computed ratio consistently yields a value of $r_{prec}(a) \approx 0.5$, a result also confirmed visually in \refFigure{fig:ratio}, which plots $r_{prec}$ versus varying $f$ for each $n$. This demonstrates that each additional fractional bit approximately halves the $\ell_2$ error.


Similarly, as $n$ increases, \refFigure{fig:l2_graph} shows that the $\ell_2$ error curves shift upward by approximately a constant factor. 
To quantify this observation, \refTable{table:l2_ratio} presents the average scaling ratio $r_{num\_q}(f)$, defined as
\begin{equation} 
	r_{num\_q}(f) = \left(\frac{L_{2}\big|_{n=16}}{L_{2}\big|_{n=8}}\right)^{\frac{1}{16-8}},
\end{equation}
which measures how quickly the error grows as $n$ increases from 8 to 16. Across all $f$'s, $r_{num\_q}(f)$ consistently remains close to 2, aligning with our theoretical prediction that each additional qubit effectively doubles the $\ell_2$ error. 
Hence, these empirical results strongly validate the scaling behavior predicted by our analytical formulation for the FP-Grover emulator.


\subsection{Equation for the Number of Fractional Bits in FP-Grover Emulation}

Having experimentally validated our error model and confirmed its scaling behavior, we next leverage these insights to guide the choice of \emph{precision} in designing the FP-Grover emulator.
Specifically, our goal is to determine the minimum number of fractional bits $f_{min}$ required to guarantee that the $\ell_2$ error remains below a desired threshold $\ell_{2,max}$.

\begin{figure}[t]
	\centering
	\includegraphics[width=0.78\textwidth]{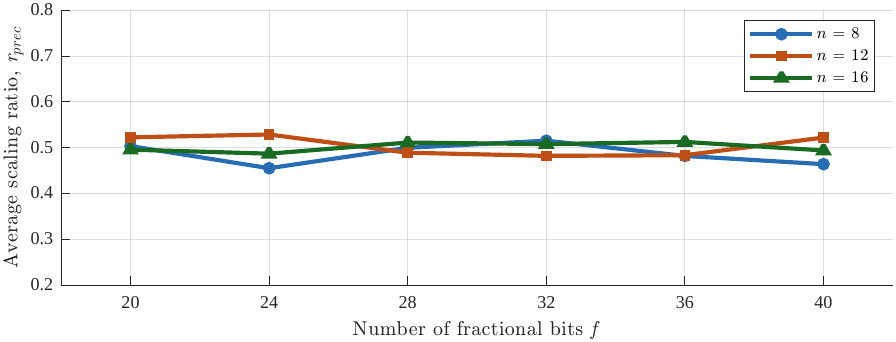}
	\caption{Average scaling ratio of the $\ell_2$ error with respect to $f$, evaluated for various qubit counts ($n=8,12,16$). }\label{fig:ratio}
\end{figure}

\paragraph*{Precision Formula.}
From Eq.~\eqref{eq:L2_big_O}, we assume
\begin{equation}
\label{eq:l2_a_n_f}
\ell_{2}
\;=\;
a\cdot2^{\,n-f},
\end{equation}
where $a$ is a proportionality constant. To pin down $a$, we substitute specific values of $(n,f,\ell_{2})$ from the midpoint of our empirical dataset in Section~\ref{subsec:validation}, namely $n=12$, $f=28$, $\ell_{2}=7.492 \times 10^{-6}$. Solving for $a$ yields $a=2^{-1.03}$, so
\begin{align}
\label{eq:l2_n_f_0933}
\ell_{2}
&=\;
2^{\,n-f-1.03},
\\
\label{eq:f_n_log2_L2_0933}
\text{or equivalently,}
\quad
f
&=\;
n
\;-\;
\log_{2}\ell_{2}
\;-\;
1.03.
\end{align}
Given a desired maximum error $\ell_{2,max}$, the minimal integer $f_{min}$ becomes
\begin{equation}
\label{eq:f_min_n_L2_max}
f_{min}
\;=\;
\Bigl\lceil 
\,n
\;-\;
\log_{2}\ell_{2,max}
\;-\;
1.03
\Bigr\rceil.
\end{equation}

\paragraph*{Experimental Validation of the Formula.}
To confirm that Eq.~\eqref{eq:f_min_n_L2_max} indeed provides a valid design rule, we tested it with $\ell_{2,max}=10^{-3},10^{-5},10^{-7}$ for various $n$. 
As reported in \refTable{table:fmin}, each calculated $f_{min}$ yielded an emulator error below $\ell_{2,max}$ when plugged back into Eq.~\eqref{eq:L2_theory}. Thus, the proposed formula is practically effective for determining the minimum fractional bits needed to meet an error specification.

\begin{table}
    \centering
    \renewcommand{\arraystretch}{1.2}
    \setlength{\tabcolsep}{6pt}
    \begin{tabular}{c|cc|cc|cc}
        \toprule
        \multirow{2}{*}{$n$} & \multicolumn{2}{c|}{$\ell_{2,max} = 10^{-3}$} & \multicolumn{2}{c|}{$\ell_{2,max} = 10^{-5}$} & \multicolumn{2}{c}{$\ell_{2,max} = 10^{-7}$} \\
        \cline{2-7}
        & $f_{min}$ & $\ell_{2}$ & $f_{min}$ & $\ell_{2}$ & $f_{min}$ & $\ell_{2}$  \\
        \midrule
        8  & $17$ & \cellcolor{orange!10} $8.271 \times 10^{-4}$  & $24$ & \cellcolor{orange!10} $4.459 \times 10^{-6}$  & $31$ & \cellcolor{orange!10} $3.825 \times 10^{-8}$  \\
        12 & $21$ & \cellcolor{orange!10} $8.742 \times 10^{-4}$  & $28$ & \cellcolor{orange!10} $7.493 \times 10^{-6}$  & $35$ & \cellcolor{orange!10} $4.354 \times 10^{-8}$ \\
        16 & $25$ & \cellcolor{orange!10} $8.778 \times 10^{-4}$  & $32$ & \cellcolor{orange!10} $7.599 \times 10^{-6}$  & $39$ & \cellcolor{orange!10} $5.859 \times 10^{-8}$  \\
        \bottomrule
    \end{tabular}
\caption{Minimal required fractional bit precision ($f_{min}$) calculated using the proposed formula for different error thresholds ($\ell_{2,max}$), along with the corresponding verification of actual $\ell_2$ errors from the FP-Grover emulator.}
    \label{table:fmin}
\end{table}

\paragraph*{Practical Design Implications.}
With Eq.~\eqref{eq:f_min_n_L2_max}, a designer can quantitatively balance \emph{accuracy} and \emph{resource efficiency} when configuring a FP-Grover emulator. 
By precomputing $f_{min}$, one avoids the memory and computational overhead of using unnecessarily high precision, yet still guarantees that the error remains below $\ell_{2,max}$. In this sense, our method provides a systematic guideline for hardware designers and system architects to choose the optimal precision level for their target error tolerance. Consequently, this study not only analyzes classical errors that arise in quantum algorithm emulation but also offers a practical resource-allocation framework for real-world FP-Grover emulator implementations.

\section{Conclusion}
\label{sec:conclusion}

In this work, we systematically investigated the impact of classical truncation errors on the accuracy of fixed-point arithmetic emulation for Grover’s quantum search algorithm. 
By introducing a simplified yet rigorously justified representation of quantum states throughout Grover’s iterations, we derived precise mathematical expressions to characterize how truncation errors propagate and accumulate across successive quantum gate operations.
Our theoretical analysis demonstrated that truncation-induced errors in the measurement probability distributions scale exponentially as $O(2^{n - f})$, where $n$ denotes the number of qubits and $f$ represents the fractional-bit precision. 
We validated this theoretical prediction comprehensively through both numerical analysis and empirical experiments using an actual fixed-point QC emulator implementation. 
These empirical validations aligned precisely with our derived theoretical results, confirming the accuracy and robustness of our analytical framework.
Crucially, our analysis provided a practical closed-form formula for selecting the minimal fractional-bit precision required to meet any specified error tolerance. 
The reliability and applicability of this formula were verified empirically, thereby offering QC emulator designers a systematic and theoretically grounded approach to balancing computational accuracy and hardware resource constraints. 
Ultimately, our work bridges theoretical error analysis and practical emulator design, laying a robust foundation for precision-aware fixed-point quantum emulation of Grover’s algorithm.

\section*{CRediT authorship contribution statement}
\textbf{Seonghyun Choi:} Resources, Data curation, Writing - Review \& Editing.
\textbf{Kyeongwon Lee:} Conceptualization, Methodology, Software, Formal analysis, Writing - Original Draft. 
\textbf{Jongin Choi:} Writing - Review \& Editing, Validation, Visualization.
\textbf{Woojoo Lee:} Conceptualization, Investigation, Writing - Original Draft \& Editing, Supervision, Funding acquisition.

\section*{Declaration of Competing Interest}
The authors have no conflicts of interest to declare.

\section*{Acknowledgements}
This work was supported in part by Institute of Information \& communications Technology Planning \& Evaluation (IITP) grant funded by the Korea government (MSIT) (No. RS-2023-00277060, Development of open edge AI SoC hardware and software platform), in part by the National Research Foundation of Korea (NRF) grant funded by the Korea government (MSIT) (No. RS-2024-00345668), and in part by the Chung-Ang University Graduate Research Scholarship Grants in 2024.

\printcredits

\bibliographystyle{elsarticle-num-names}

\bibliography{references}

\begin{thebibliography}{41}
\expandafter\ifx\csname natexlab\endcsname\relax\def\natexlab#1{#1}\fi
\providecommand{\url}[1]{\texttt{#1}}
\providecommand{\href}[2]{#2}
\providecommand{\path}[1]{#1}
\providecommand{\DOIprefix}{doi:}
\providecommand{\ArXivprefix}{arXiv:}
\providecommand{\URLprefix}{URL: }
\providecommand{\Pubmedprefix}{pmid:}
\providecommand{\doi}[1]{\href{http://dx.doi.org/#1}{\path{#1}}}
\providecommand{\Pubmed}[1]{\href{pmid:#1}{\path{#1}}}
\providecommand{\bibinfo}[2]{#2}
\ifx\xfnm\relax \def\xfnm[#1]{\unskip,\space#1}\fi
\bibitem[{Kim et~al.(2023)Kim, Eddins, Anand, Wei, Van Den~Berg, Rosenblatt,
  Nayfeh, Wu, Zaletel, Temme et~al.}]{Kim:Nature23}
\bibinfo{author}{Y.~Kim}, \bibinfo{author}{A.~Eddins},
  \bibinfo{author}{S.~Anand}, \bibinfo{author}{K.~X. Wei},
  \bibinfo{author}{E.~Van Den~Berg}, \bibinfo{author}{S.~Rosenblatt},
  \bibinfo{author}{H.~Nayfeh}, \bibinfo{author}{Y.~Wu},
  \bibinfo{author}{M.~Zaletel}, \bibinfo{author}{K.~Temme}, et~al.,
\newblock \bibinfo{title}{Evidence for the utility of quantum computing before
  fault tolerance},
\newblock \bibinfo{journal}{Nature} \bibinfo{volume}{618}
  (\bibinfo{year}{2023}) \bibinfo{pages}{500--505}.
\bibitem[{Kikuchi et~al.(2023)Kikuchi, Mc~Keever, Coopmans, Lubasch, and
  Benedetti}]{Kikuchi:Nature23}
\bibinfo{author}{Y.~Kikuchi}, \bibinfo{author}{C.~Mc~Keever},
  \bibinfo{author}{L.~Coopmans}, \bibinfo{author}{M.~Lubasch},
  \bibinfo{author}{M.~Benedetti},
\newblock \bibinfo{title}{Realization of quantum signal processing on a noisy
  quantum computer},
\newblock \bibinfo{journal}{npj Quantum Information} \bibinfo{volume}{9}
  (\bibinfo{year}{2023}) \bibinfo{pages}{93}.
\bibitem[{{Google AI Quantum} et~al.(2020){Google AI Quantum},
  Collaborators*†, Arute, Arya, Babbush, Bacon, Bardin, Barends, Boixo,
  Broughton, Buckley, Buell, Burkett, Bushnell, Chen, Chen, Chiaro, Collins,
  Courtney, Demura, Dunsworth, Farhi, Fowler, Foxen, Gidney, Giustina, Graff,
  Habegger, Harrigan, Ho, Hong, Huang, Huggins, Ioffe, Isakov, Jeffrey, Jiang,
  Jones, Kafri, Kechedzhi, Kelly, Kim, Klimov, Korotkov, Kostritsa, Landhuis,
  Laptev, Lindmark, Lucero, Martin, Martinis, McClean, McEwen, Megrant, Mi,
  Mohseni, Mruczkiewicz, Mutus, Naaman, Neeley, Neill, Neven, Niu, O’Brien,
  Ostby, Petukhov, Putterman, Quintana, Roushan, Rubin, Sank, Satzinger,
  Smelyanskiy, Strain, Sung, Szalay, Takeshita, Vainsencher, White, Wiebe, Yao,
  Yeh, and Zalcman}]{Google:Science20}
\bibinfo{author}{{Google AI Quantum}}, \bibinfo{author}{Collaborators*†},
  \bibinfo{author}{F.~Arute}, \bibinfo{author}{K.~Arya},
  \bibinfo{author}{R.~Babbush}, \bibinfo{author}{D.~Bacon},
  \bibinfo{author}{J.~C. Bardin}, \bibinfo{author}{R.~Barends},
  \bibinfo{author}{S.~Boixo}, \bibinfo{author}{M.~Broughton},
  \bibinfo{author}{B.~B. Buckley}, \bibinfo{author}{D.~A. Buell},
  \bibinfo{author}{B.~Burkett}, \bibinfo{author}{N.~Bushnell},
  \bibinfo{author}{Y.~Chen}, \bibinfo{author}{Z.~Chen},
  \bibinfo{author}{B.~Chiaro}, \bibinfo{author}{R.~Collins},
  \bibinfo{author}{W.~Courtney}, \bibinfo{author}{S.~Demura},
  \bibinfo{author}{A.~Dunsworth}, \bibinfo{author}{E.~Farhi},
  \bibinfo{author}{A.~Fowler}, \bibinfo{author}{B.~Foxen},
  \bibinfo{author}{C.~Gidney}, \bibinfo{author}{M.~Giustina},
  \bibinfo{author}{R.~Graff}, \bibinfo{author}{S.~Habegger},
  \bibinfo{author}{M.~P. Harrigan}, \bibinfo{author}{A.~Ho},
  \bibinfo{author}{S.~Hong}, \bibinfo{author}{T.~Huang}, \bibinfo{author}{W.~J.
  Huggins}, \bibinfo{author}{L.~Ioffe}, \bibinfo{author}{S.~V. Isakov},
  \bibinfo{author}{E.~Jeffrey}, \bibinfo{author}{Z.~Jiang},
  \bibinfo{author}{C.~Jones}, \bibinfo{author}{D.~Kafri},
  \bibinfo{author}{K.~Kechedzhi}, \bibinfo{author}{J.~Kelly},
  \bibinfo{author}{S.~Kim}, \bibinfo{author}{P.~V. Klimov},
  \bibinfo{author}{A.~Korotkov}, \bibinfo{author}{F.~Kostritsa},
  \bibinfo{author}{D.~Landhuis}, \bibinfo{author}{P.~Laptev},
  \bibinfo{author}{M.~Lindmark}, \bibinfo{author}{E.~Lucero},
  \bibinfo{author}{O.~Martin}, \bibinfo{author}{J.~M. Martinis},
  \bibinfo{author}{J.~R. McClean}, \bibinfo{author}{M.~McEwen},
  \bibinfo{author}{A.~Megrant}, \bibinfo{author}{X.~Mi},
  \bibinfo{author}{M.~Mohseni}, \bibinfo{author}{W.~Mruczkiewicz},
  \bibinfo{author}{J.~Mutus}, \bibinfo{author}{O.~Naaman},
  \bibinfo{author}{M.~Neeley}, \bibinfo{author}{C.~Neill},
  \bibinfo{author}{H.~Neven}, \bibinfo{author}{M.~Y. Niu},
  \bibinfo{author}{T.~E. O’Brien}, \bibinfo{author}{E.~Ostby},
  \bibinfo{author}{A.~Petukhov}, \bibinfo{author}{H.~Putterman},
  \bibinfo{author}{C.~Quintana}, \bibinfo{author}{P.~Roushan},
  \bibinfo{author}{N.~C. Rubin}, \bibinfo{author}{D.~Sank},
  \bibinfo{author}{K.~J. Satzinger}, \bibinfo{author}{V.~Smelyanskiy},
  \bibinfo{author}{D.~Strain}, \bibinfo{author}{K.~J. Sung},
  \bibinfo{author}{M.~Szalay}, \bibinfo{author}{T.~Y. Takeshita},
  \bibinfo{author}{A.~Vainsencher}, \bibinfo{author}{T.~White},
  \bibinfo{author}{N.~Wiebe}, \bibinfo{author}{Z.~J. Yao},
  \bibinfo{author}{P.~Yeh}, \bibinfo{author}{A.~Zalcman},
\newblock \bibinfo{title}{Hartree-fock on a superconducting qubit quantum
  computer},
\newblock \bibinfo{journal}{Science} \bibinfo{volume}{369}
  (\bibinfo{year}{2020}) \bibinfo{pages}{1084--1089}.
\bibitem[{Zhang et~al.(2024)Zhang, Boothby, and Kamenev}]{Zhang:Nature24}
\bibinfo{author}{H.~Zhang}, \bibinfo{author}{K.~Boothby},
  \bibinfo{author}{A.~Kamenev},
\newblock \bibinfo{title}{Cyclic quantum annealing: searching for deep
  low-energy states in 5000-qubit spin glass},
\newblock \bibinfo{journal}{Scientific Reports} \bibinfo{volume}{14}
  (\bibinfo{year}{2024}) \bibinfo{pages}{30784}.
\bibitem[{{Microsoft Azure Quantum} et~al.(2025){Microsoft Azure Quantum},
  Aghaee, Alcaraz~Ramirez, Alam, Ali, Andrzejczuk, Antipov, Astafev, Barzegar,
  Bauer et~al.}]{Microsoft:Nature25}
\bibinfo{author}{{Microsoft Azure Quantum}}, \bibinfo{author}{M.~Aghaee},
  \bibinfo{author}{A.~Alcaraz~Ramirez}, \bibinfo{author}{Z.~Alam},
  \bibinfo{author}{R.~Ali}, \bibinfo{author}{M.~Andrzejczuk},
  \bibinfo{author}{A.~Antipov}, \bibinfo{author}{M.~Astafev},
  \bibinfo{author}{A.~Barzegar}, \bibinfo{author}{B.~Bauer}, et~al.,
\newblock \bibinfo{title}{Interferometric single-shot parity measurement in
  inas--al hybrid devices},
\newblock \bibinfo{journal}{Nature} \bibinfo{volume}{638}
  (\bibinfo{year}{2025}) \bibinfo{pages}{651--655}.
\bibitem[{Ladd et~al.(2010)Ladd, Jelezko, Laflamme, Nakamura, Monroe, and
  O’Brien}]{Ladd:Nature10}
\bibinfo{author}{T.~D. Ladd}, \bibinfo{author}{F.~Jelezko},
  \bibinfo{author}{R.~Laflamme}, \bibinfo{author}{Y.~Nakamura},
  \bibinfo{author}{C.~Monroe}, \bibinfo{author}{J.~L. O’Brien},
\newblock \bibinfo{title}{Quantum computers},
\newblock \bibinfo{journal}{nature} \bibinfo{volume}{464}
  (\bibinfo{year}{2010}) \bibinfo{pages}{45--53}.
\bibitem[{Cao et~al.(2019)Cao, Romero, Olson, Degroote, Johnson, Kieferov{\'a},
  Kivlichan, Menke, Peropadre, Sawaya et~al.}]{Cao:ACS19}
\bibinfo{author}{Y.~Cao}, \bibinfo{author}{J.~Romero}, \bibinfo{author}{J.~P.
  Olson}, \bibinfo{author}{M.~Degroote}, \bibinfo{author}{P.~D. Johnson},
  \bibinfo{author}{M.~Kieferov{\'a}}, \bibinfo{author}{I.~D. Kivlichan},
  \bibinfo{author}{T.~Menke}, \bibinfo{author}{B.~Peropadre},
  \bibinfo{author}{N.~P. Sawaya}, et~al.,
\newblock \bibinfo{title}{Quantum chemistry in the age of quantum computing},
\newblock \bibinfo{journal}{Chemical reviews} \bibinfo{volume}{119}
  (\bibinfo{year}{2019}) \bibinfo{pages}{10856--10915}.
\bibitem[{Gyongyosi and Imre(2019)}]{Gyongyosi:Elsevier19}
\bibinfo{author}{L.~Gyongyosi}, \bibinfo{author}{S.~Imre},
\newblock \bibinfo{title}{A survey on quantum computing technology},
\newblock \bibinfo{journal}{Computer Science Review} \bibinfo{volume}{31}
  (\bibinfo{year}{2019}) \bibinfo{pages}{51--71}.
\bibitem[{Arute et~al.(2019)Arute, Arya, Babbush, Bacon, Bardin, Barends,
  Biswas, Boixo, Brandao, Buell et~al.}]{Arute:Nature19}
\bibinfo{author}{F.~Arute}, \bibinfo{author}{K.~Arya},
  \bibinfo{author}{R.~Babbush}, \bibinfo{author}{D.~Bacon},
  \bibinfo{author}{J.~C. Bardin}, \bibinfo{author}{R.~Barends},
  \bibinfo{author}{R.~Biswas}, \bibinfo{author}{S.~Boixo},
  \bibinfo{author}{F.~G. Brandao}, \bibinfo{author}{D.~A. Buell}, et~al.,
\newblock \bibinfo{title}{Quantum supremacy using a programmable
  superconducting processor},
\newblock \bibinfo{journal}{Nature} \bibinfo{volume}{574}
  (\bibinfo{year}{2019}) \bibinfo{pages}{505--510}.
\bibitem[{Zhong et~al.(2020)Zhong, Wang, Deng, Chen, Peng, Luo, Qin, Wu, Ding,
  Hu et~al.}]{Zhong:Science20}
\bibinfo{author}{H.-S. Zhong}, \bibinfo{author}{H.~Wang},
  \bibinfo{author}{Y.-H. Deng}, \bibinfo{author}{M.-C. Chen},
  \bibinfo{author}{L.-C. Peng}, \bibinfo{author}{Y.-H. Luo},
  \bibinfo{author}{J.~Qin}, \bibinfo{author}{D.~Wu}, \bibinfo{author}{X.~Ding},
  \bibinfo{author}{Y.~Hu}, et~al.,
\newblock \bibinfo{title}{Quantum computational advantage using photons},
\newblock \bibinfo{journal}{Science} \bibinfo{volume}{370}
  (\bibinfo{year}{2020}) \bibinfo{pages}{1460--1463}.
\bibitem[{Kelly et~al.(2015)Kelly, Barends, Fowler, Megrant, Jeffrey, White,
  Sank, Mutus, Campbell, Chen et~al.}]{Kelly:Nature15}
\bibinfo{author}{J.~Kelly}, \bibinfo{author}{R.~Barends},
  \bibinfo{author}{A.~G. Fowler}, \bibinfo{author}{A.~Megrant},
  \bibinfo{author}{E.~Jeffrey}, \bibinfo{author}{T.~C. White},
  \bibinfo{author}{D.~Sank}, \bibinfo{author}{J.~Y. Mutus},
  \bibinfo{author}{B.~Campbell}, \bibinfo{author}{Y.~Chen}, et~al.,
\newblock \bibinfo{title}{State preservation by repetitive error detection in a
  superconducting quantum circuit},
\newblock \bibinfo{journal}{Nature} \bibinfo{volume}{519}
  (\bibinfo{year}{2015}) \bibinfo{pages}{66--69}.
\bibitem[{Maurand et~al.(2016)Maurand, Jehl, Kotekar-Patil, Corna,
  Bohuslavskyi, Lavi{\'e}ville, Hutin, Barraud, Vinet, Sanquer
  et~al.}]{Maurand:Nature16}
\bibinfo{author}{R.~Maurand}, \bibinfo{author}{X.~Jehl},
  \bibinfo{author}{D.~Kotekar-Patil}, \bibinfo{author}{A.~Corna},
  \bibinfo{author}{H.~Bohuslavskyi}, \bibinfo{author}{R.~Lavi{\'e}ville},
  \bibinfo{author}{L.~Hutin}, \bibinfo{author}{S.~Barraud},
  \bibinfo{author}{M.~Vinet}, \bibinfo{author}{M.~Sanquer}, et~al.,
\newblock \bibinfo{title}{A cmos silicon spin qubit},
\newblock \bibinfo{journal}{Nature communications} \bibinfo{volume}{7}
  (\bibinfo{year}{2016}) \bibinfo{pages}{13575}.
\bibitem[{Friis et~al.(2018)Friis, Marty, Maier, Hempel, Holz{\"a}pfel,
  Jurcevic, Plenio, Huber, Roos, Blatt et~al.}]{Friis:APS18}
\bibinfo{author}{N.~Friis}, \bibinfo{author}{O.~Marty},
  \bibinfo{author}{C.~Maier}, \bibinfo{author}{C.~Hempel},
  \bibinfo{author}{M.~Holz{\"a}pfel}, \bibinfo{author}{P.~Jurcevic},
  \bibinfo{author}{M.~B. Plenio}, \bibinfo{author}{M.~Huber},
  \bibinfo{author}{C.~Roos}, \bibinfo{author}{R.~Blatt}, et~al.,
\newblock \bibinfo{title}{Observation of entangled states of a fully controlled
  20-qubit system},
\newblock \bibinfo{journal}{Physical Review X} \bibinfo{volume}{8}
  (\bibinfo{year}{2018}) \bibinfo{pages}{021012}.
\bibitem[{Ajagekar and You(2019)}]{Ajagekar:Elsevier19}
\bibinfo{author}{A.~Ajagekar}, \bibinfo{author}{F.~You},
\newblock \bibinfo{title}{Quantum computing for energy systems optimization:
  Challenges and opportunities},
\newblock \bibinfo{journal}{Energy} \bibinfo{volume}{179}
  (\bibinfo{year}{2019}) \bibinfo{pages}{76--89}.
\bibitem[{Zhou et~al.(2020)Zhou, Wang, Choi, Pichler, and Lukin}]{Zhou:PRX20}
\bibinfo{author}{L.~Zhou}, \bibinfo{author}{S.-T. Wang},
  \bibinfo{author}{S.~Choi}, \bibinfo{author}{H.~Pichler},
  \bibinfo{author}{M.~D. Lukin},
\newblock \bibinfo{title}{Quantum approximate optimization algorithm:
  Performance, mechanism, and implementation on near-term devices},
\newblock \bibinfo{journal}{Physical Review X} \bibinfo{volume}{10}
  (\bibinfo{year}{2020}) \bibinfo{pages}{021067}.
\bibitem[{Wang et~al.(2023)Wang, Kim, and Suresh}]{Wang:ASME23}
\bibinfo{author}{Y.~Wang}, \bibinfo{author}{J.~E. Kim},
  \bibinfo{author}{K.~Suresh},
\newblock \bibinfo{title}{Opportunities and challenges of quantum computing for
  engineering optimization},
\newblock \bibinfo{journal}{Journal of Computing and Information Science in
  Engineering} \bibinfo{volume}{23} (\bibinfo{year}{2023})
  \bibinfo{pages}{060817}.
\bibitem[{Cai et~al.(2015)Cai, Wu, Su, Chen, Wang, Li, Liu, Lu, and
  Pan}]{Cai:APS15}
\bibinfo{author}{X.-D. Cai}, \bibinfo{author}{D.~Wu}, \bibinfo{author}{Z.-E.
  Su}, \bibinfo{author}{M.-C. Chen}, \bibinfo{author}{X.-L. Wang},
  \bibinfo{author}{L.~Li}, \bibinfo{author}{N.-L. Liu}, \bibinfo{author}{C.-Y.
  Lu}, \bibinfo{author}{J.-W. Pan},
\newblock \bibinfo{title}{Entanglement-based machine learning on a quantum
  computer},
\newblock \bibinfo{journal}{Physical review letters} \bibinfo{volume}{114}
  (\bibinfo{year}{2015}) \bibinfo{pages}{110504}.
\bibitem[{Ramezani et~al.(2020)Ramezani, Sommers, Manchukonda, Rahimi, and
  Amirlatifi}]{Ramezani:IEEE20}
\bibinfo{author}{S.~B. Ramezani}, \bibinfo{author}{A.~Sommers},
  \bibinfo{author}{H.~K. Manchukonda}, \bibinfo{author}{S.~Rahimi},
  \bibinfo{author}{A.~Amirlatifi},
\newblock \bibinfo{title}{Machine learning algorithms in quantum computing: A
  survey},
\newblock in: \bibinfo{booktitle}{2020 International joint conference on neural
  networks (IJCNN)}, \bibinfo{organization}{IEEE}, \bibinfo{year}{2020}, pp.
  \bibinfo{pages}{1--8}.
\bibitem[{Lloyd et~al.(2013)Lloyd, Mohseni, and Rebentrost}]{Lloyd:arxiv13}
\bibinfo{author}{S.~Lloyd}, \bibinfo{author}{M.~Mohseni},
  \bibinfo{author}{P.~Rebentrost},
\newblock \bibinfo{title}{Quantum algorithms for supervised and unsupervised
  machine learning},
\newblock \bibinfo{journal}{arXiv preprint arXiv:1307.0411}
  (\bibinfo{year}{2013}).
\bibitem[{Iriyama and Ohya(2012)}]{Iriyama:AMC12}
\bibinfo{author}{S.~Iriyama}, \bibinfo{author}{M.~Ohya},
\newblock \bibinfo{title}{Computational complexity and applications of quantum
  algorithm},
\newblock \bibinfo{journal}{Applied Mathematics and Computation}
  \bibinfo{volume}{218} (\bibinfo{year}{2012}) \bibinfo{pages}{8019--8028}.
\bibitem[{Ollitrault et~al.(2021)Ollitrault, Miessen, and
  Tavernelli}]{Ollitrault:ACS21}
\bibinfo{author}{P.~J. Ollitrault}, \bibinfo{author}{A.~Miessen},
  \bibinfo{author}{I.~Tavernelli},
\newblock \bibinfo{title}{Molecular quantum dynamics: A quantum computing
  perspective},
\newblock \bibinfo{journal}{Accounts of Chemical Research} \bibinfo{volume}{54}
  (\bibinfo{year}{2021}) \bibinfo{pages}{4229--4238}.
\bibitem[{Outeiral et~al.(2021)Outeiral, Strahm, Shi, Morris, Benjamin, and
  Deane}]{Outeiral:wiley21}
\bibinfo{author}{C.~Outeiral}, \bibinfo{author}{M.~Strahm},
  \bibinfo{author}{J.~Shi}, \bibinfo{author}{G.~M. Morris},
  \bibinfo{author}{S.~C. Benjamin}, \bibinfo{author}{C.~M. Deane},
\newblock \bibinfo{title}{The prospects of quantum computing in computational
  molecular biology},
\newblock \bibinfo{journal}{Wiley Interdisciplinary Reviews: Computational
  Molecular Science} \bibinfo{volume}{11} (\bibinfo{year}{2021})
  \bibinfo{pages}{e1481}.
\bibitem[{Buluta and Nori(2009)}]{Buluta:Science09}
\bibinfo{author}{I.~Buluta}, \bibinfo{author}{F.~Nori},
\newblock \bibinfo{title}{Quantum simulators},
\newblock \bibinfo{journal}{Science} \bibinfo{volume}{326}
  (\bibinfo{year}{2009}) \bibinfo{pages}{108--111}.
\bibitem[{H{\"a}ner et~al.(2016)H{\"a}ner, Steiger, Smelyanskiy, and
  Troyer}]{Haner:IEEE16}
\bibinfo{author}{T.~H{\"a}ner}, \bibinfo{author}{D.~S. Steiger},
  \bibinfo{author}{M.~Smelyanskiy}, \bibinfo{author}{M.~Troyer},
\newblock \bibinfo{title}{High performance emulation of quantum circuits},
\newblock in: \bibinfo{booktitle}{SC'16: Proceedings of the International
  Conference for High Performance Computing, Networking, Storage and Analysis},
  \bibinfo{organization}{IEEE}, \bibinfo{year}{2016}, pp.
  \bibinfo{pages}{866--874}.
\bibitem[{Li and Pang(2021)}]{Li:IEEE21}
\bibinfo{author}{H.~Li}, \bibinfo{author}{Y.~Pang},
\newblock \bibinfo{title}{Fpga-accelerated quantum computing emulation and
  quantum key distillation},
\newblock \bibinfo{journal}{IEEE Micro} \bibinfo{volume}{41}
  (\bibinfo{year}{2021}) \bibinfo{pages}{49--57}.
\bibitem[{Altman et~al.(2021)Altman, Brown, Carleo, Carr, Demler, Chin,
  DeMarco, Economou, Eriksson, Fu, Greiner, Hazzard, Hulet, Koll\'ar, Lev,
  Lukin, Ma, Mi, Misra, Monroe, Murch, Nazario, Ni, Potter, Roushan, Saffman,
  Schleier-Smith, Siddiqi, Simmonds, Singh, Spielman, Temme, Weiss, Vu\ifmmode
  \check{c}\else \v{c}\fi{}kovi\ifmmode~\acute{c}\else \'{c}\fi{},
  Vuleti\ifmmode~\acute{c}\else \'{c}\fi{}, Ye, and Zwierlein}]{Altman:PRX21}
\bibinfo{author}{E.~Altman}, \bibinfo{author}{K.~R. Brown},
  \bibinfo{author}{G.~Carleo}, \bibinfo{author}{L.~D. Carr},
  \bibinfo{author}{E.~Demler}, \bibinfo{author}{C.~Chin},
  \bibinfo{author}{B.~DeMarco}, \bibinfo{author}{S.~E. Economou},
  \bibinfo{author}{M.~A. Eriksson}, \bibinfo{author}{K.-M.~C. Fu},
  \bibinfo{author}{M.~Greiner}, \bibinfo{author}{K.~R. Hazzard},
  \bibinfo{author}{R.~G. Hulet}, \bibinfo{author}{A.~J. Koll\'ar},
  \bibinfo{author}{B.~L. Lev}, \bibinfo{author}{M.~D. Lukin},
  \bibinfo{author}{R.~Ma}, \bibinfo{author}{X.~Mi}, \bibinfo{author}{S.~Misra},
  \bibinfo{author}{C.~Monroe}, \bibinfo{author}{K.~Murch},
  \bibinfo{author}{Z.~Nazario}, \bibinfo{author}{K.-K. Ni},
  \bibinfo{author}{A.~C. Potter}, \bibinfo{author}{P.~Roushan},
  \bibinfo{author}{M.~Saffman}, \bibinfo{author}{M.~Schleier-Smith},
  \bibinfo{author}{I.~Siddiqi}, \bibinfo{author}{R.~Simmonds},
  \bibinfo{author}{M.~Singh}, \bibinfo{author}{I.~Spielman},
  \bibinfo{author}{K.~Temme}, \bibinfo{author}{D.~S. Weiss},
  \bibinfo{author}{J.~Vu\ifmmode \check{c}\else
  \v{c}\fi{}kovi\ifmmode~\acute{c}\else \'{c}\fi{}},
  \bibinfo{author}{V.~Vuleti\ifmmode~\acute{c}\else \'{c}\fi{}},
  \bibinfo{author}{J.~Ye}, \bibinfo{author}{M.~Zwierlein},
\newblock \bibinfo{title}{Quantum simulators: Architectures and opportunities},
\newblock \bibinfo{journal}{PRX Quantum} \bibinfo{volume}{2}
  (\bibinfo{year}{2021}) \bibinfo{pages}{017003}.
\bibitem[{Choi and Lee(2024)}]{Choi:AIMS24}
\bibinfo{author}{S.~Choi}, \bibinfo{author}{W.~Lee},
\newblock \bibinfo{title}{Developing a grover’s quantum algorithm emulator on
  standalone fpgas: optimization and implementation},
\newblock \bibinfo{journal}{AIMS Mathematics} \bibinfo{volume}{9}
  (\bibinfo{year}{2024}) \bibinfo{pages}{30939--30971}.
\bibitem[{Choi et~al.(2025)Choi, Lee, Lee, and Lee}]{Choi:arXiv25}
\bibinfo{author}{S.~Choi}, \bibinfo{author}{K.~Lee}, \bibinfo{author}{J.-J.
  Lee}, \bibinfo{author}{W.~Lee}, \bibinfo{title}{Standalone fpga-based qaoa
  emulator for weighted-maxcut on embedded devices}, \bibinfo{year}{2025}.
  \URLprefix \url{https://arxiv.org/abs/2502.11316}.
  \href{http://arxiv.org/abs/2502.11316}{{\tt arXiv:2502.11316}}.
\bibitem[{Khalid et~al.(2004)Khalid, Zilic, and Radecka}]{Khalid:IEEE04}
\bibinfo{author}{A.~U. Khalid}, \bibinfo{author}{Z.~Zilic},
  \bibinfo{author}{K.~Radecka},
\newblock \bibinfo{title}{Fpga emulation of quantum circuits},
\newblock in: \bibinfo{booktitle}{IEEE International Conference on Computer
  Design: VLSI in Computers and Processors, 2004. ICCD 2004. Proceedings.},
  \bibinfo{organization}{IEEE}, \bibinfo{year}{2004}, pp.
  \bibinfo{pages}{310--315}.
\bibitem[{Grover(1996)}]{Grover:ACM96}
\bibinfo{author}{L.~K. Grover},
\newblock \bibinfo{title}{A fast quantum mechanical algorithm for database
  search},
\newblock in: \bibinfo{booktitle}{Proceedings of the twenty-eighth annual ACM
  symposium on Theory of computing}, \bibinfo{year}{1996}, pp.
  \bibinfo{pages}{212--219}.
\bibitem[{Bag et~al.(2022)Bag, Goswami, and Kandpal}]{Bag:IEEE22}
\bibinfo{author}{K.~Bag}, \bibinfo{author}{M.~Goswami},
  \bibinfo{author}{K.~Kandpal},
\newblock \bibinfo{title}{Fpga based resource efficient simulation and
  emulation of grover’s search algorithm},
\newblock in: \bibinfo{booktitle}{2022 IEEE 19th India Council International
  Conference (INDICON)}, \bibinfo{organization}{IEEE}, \bibinfo{year}{2022},
  pp. \bibinfo{pages}{1--6}.
\bibitem[{Bennett et~al.(1997)Bennett, Bernstein, Brassard, and
  Vazirani}]{Bennett:SIAM97}
\bibinfo{author}{C.~H. Bennett}, \bibinfo{author}{E.~Bernstein},
  \bibinfo{author}{G.~Brassard}, \bibinfo{author}{U.~Vazirani},
\newblock \bibinfo{title}{Strengths and weaknesses of quantum computing},
\newblock \bibinfo{journal}{SIAM journal on Computing} \bibinfo{volume}{26}
  (\bibinfo{year}{1997}) \bibinfo{pages}{1510--1523}.
\bibitem[{Boyer et~al.(1998)Boyer, Brassard, H{\o}yer, and
  Tapp}]{Boyer:Wiley98}
\bibinfo{author}{M.~Boyer}, \bibinfo{author}{G.~Brassard},
  \bibinfo{author}{P.~H{\o}yer}, \bibinfo{author}{A.~Tapp},
\newblock \bibinfo{title}{Tight bounds on quantum searching},
\newblock \bibinfo{journal}{Fortschritte der Physik: Progress of Physics}
  \bibinfo{volume}{46} (\bibinfo{year}{1998}) \bibinfo{pages}{493--505}.
\bibitem[{Zalka(1999)}]{Zalka:PRA99}
\bibinfo{author}{C.~Zalka},
\newblock \bibinfo{title}{Grover’s quantum searching algorithm is optimal},
\newblock \bibinfo{journal}{Physical Review A} \bibinfo{volume}{60}
  (\bibinfo{year}{1999}) \bibinfo{pages}{2746}.
\bibitem[{Preston(2022)}]{Preston:IEEE22}
\bibinfo{author}{R.~H. Preston},
\newblock \bibinfo{title}{Applying grover's algorithm to hash functions: a
  software perspective},
\newblock \bibinfo{journal}{IEEE transactions on quantum engineering}
  \bibinfo{volume}{3} (\bibinfo{year}{2022}) \bibinfo{pages}{1--10}.
\bibitem[{Grassl et~al.(2016)Grassl, Langenberg, Roetteler, and
  Steinwandt}]{Grassl:Springer16}
\bibinfo{author}{M.~Grassl}, \bibinfo{author}{B.~Langenberg},
  \bibinfo{author}{M.~Roetteler}, \bibinfo{author}{R.~Steinwandt},
\newblock \bibinfo{title}{Applying grover’s algorithm to aes: quantum
  resource estimates},
\newblock in: \bibinfo{booktitle}{International Workshop on Post-Quantum
  Cryptography}, \bibinfo{organization}{Springer}, \bibinfo{year}{2016}, pp.
  \bibinfo{pages}{29--43}.
\bibitem[{Schwabe and Westerbaan(2016)}]{Schwabe:Springer16}
\bibinfo{author}{P.~Schwabe}, \bibinfo{author}{B.~Westerbaan},
\newblock \bibinfo{title}{Solving binary with grover’s algorithm},
\newblock in: \bibinfo{booktitle}{International Conference on Security,
  Privacy, and Applied Cryptography Engineering},
  \bibinfo{organization}{Springer}, \bibinfo{year}{2016}, pp.
  \bibinfo{pages}{303--322}.
\bibitem[{Habibi et~al.(2022)Habibi, Golestan, Soltanmanesh, Guerrero, and
  Vasquez}]{Habibi:MDPI22}
\bibinfo{author}{M.~R. Habibi}, \bibinfo{author}{S.~Golestan},
  \bibinfo{author}{A.~Soltanmanesh}, \bibinfo{author}{J.~M. Guerrero},
  \bibinfo{author}{J.~C. Vasquez},
\newblock \bibinfo{title}{Power and energy applications based on quantum
  computing: The possible potentials of grover’s algorithm},
\newblock \bibinfo{journal}{Electronics} \bibinfo{volume}{11}
  (\bibinfo{year}{2022}) \bibinfo{pages}{2919}.
\bibitem[{Bogatyrev and Moskvin(2023)}]{Bogatyrev:IEEE23application}
\bibinfo{author}{V.~Bogatyrev}, \bibinfo{author}{V.~Moskvin},
\newblock \bibinfo{title}{Application of grover’s algorithm in route
  optimization},
\newblock in: \bibinfo{booktitle}{2023 Intelligent Technologies and Electronic
  Devices in Vehicle and Road Transport Complex (TIRVED)},
  \bibinfo{organization}{IEEE}, \bibinfo{year}{2023}, pp.
  \bibinfo{pages}{1--5}.
\bibitem[{Chakrabarty et~al.(2017)Chakrabarty, Khan, and
  Singh}]{Chakrabarty:Springer17}
\bibinfo{author}{I.~Chakrabarty}, \bibinfo{author}{S.~Khan},
  \bibinfo{author}{V.~Singh},
\newblock \bibinfo{title}{Dynamic grover search: Applications in recommendation
  systems and optimization problems},
\newblock \bibinfo{journal}{Quantum Information Processing}
  \bibinfo{volume}{16} (\bibinfo{year}{2017}) \bibinfo{pages}{1--21}.
\bibitem[{Zalka(1998)}]{Zalka:royalsociety98}
\bibinfo{author}{C.~Zalka},
\newblock \bibinfo{title}{Simulating quantum systems on a quantum computer},
\newblock \bibinfo{journal}{Proceedings of the Royal Society of London. Series
  A: Mathematical, Physical and Engineering Sciences} \bibinfo{volume}{454}
  (\bibinfo{year}{1998}) \bibinfo{pages}{313--322}.

\end{thebibliography}



\end{document}